\documentclass[12pt,a4paper]{article}

\usepackage{graphicx}  

\usepackage{xspace}
\usepackage{color}
\usepackage{colortbl}

\usepackage{amsmath}
\usepackage{subfigure}
\usepackage{ifthen} 

\newboolean{pdflatex}
\setboolean{pdflatex}{false} 
\newboolean{uprightparticles}
\setboolean{uprightparticles}{false} 
\usepackage{amssymb}
\usepackage{amsfonts}
\usepackage{upgreek}
\ifthenelse{\boolean{uprightparticles}}%
{

 \def\PDelta      {\ensuremath{\Delta}\xspace}                 
 \def\PXi      {\ensuremath{\Xi}\xspace}                 
 \def\PLambda      {\ensuremath{\Lambda}\xspace}                 
 \def\PSigma      {\ensuremath{\Sigma}\xspace}                 
 \def\POmega      {\ensuremath{\Omega}\xspace}                 
 \def\PUpsilon      {\ensuremath{\Upsilon}\xspace}                 
 
 \def\PB      {\ensuremath{\mathrm{B}}\xspace}                 
                  
 \def\PD      {\ensuremath{\mathrm{D}}\xspace}

 \def\PK      {\ensuremath{\mathrm{K}}\xspace}

 \def\Pi      {\ensuremath{\mathrm{i}}\xspace}

}
{

 \mathchardef\PDelta="7101
 \mathchardef\PXi="7104
 \mathchardef\PLambda="7103
 \mathchardef\PSigma="7106
 \mathchardef\POmega="710A
 \mathchardef\PUpsilon="7107
                  
 \def\PB      {\ensuremath{B}\xspace}                 
                  
 \def\PD      {\ensuremath{D}\xspace}

 \def\PK      {\ensuremath{K}\xspace}

 \def\Pi      {\ensuremath{i}\xspace}

}

\def\kaon  {\ensuremath{\PK}\xspace}
  \def\Kbar  {\kern 0.2em\overline{\kern -0.2em \PK}{}\xspace}

\def\Kz    {\ensuremath{\kaon^0}\xspace}
\def\Kzb   {\ensuremath{\Kbar^0}\xspace}
\def\KzKzb {\ensuremath{\Kz \kern -0.16em \Kzb}\xspace}
\def\Kp    {\ensuremath{\kaon^+}\xspace}
\def\Km    {\ensuremath{\kaon^-}\xspace}

\def\KpKm  {\ensuremath{\Kp \kern -0.16em \Km}\xspace}

  \def\Dbar    {\kern 0.2em\overline{\kern -0.2em \PD}{}\xspace}
\def\D       {\ensuremath{\PD}\xspace}

\def\Dz      {\ensuremath{\D^0}\xspace}
\def\Dzb     {\ensuremath{\Dbar^0}\xspace}
\def\DzDzb   {\ensuremath{\Dz {\kern -0.16em \Dzb}}\xspace}
\def\Dp      {\ensuremath{\D^+}\xspace}
\def\Dm      {\ensuremath{\D^-}\xspace}

\def\DpDm    {\ensuremath{\Dp {\kern -0.16em \Dm}}\xspace}

  \def\Bbar    {\kern 0.18em\overline{\kern -0.18em \PB}{}\xspace}

  \def\Y#1S{\ensuremath{\PUpsilon{(#1S)}}\xspace}

\newcommand{\tev}{\ensuremath{\mathrm{\,Te\kern -0.1em V}}\xspace}
\newcommand{\gev}{\ensuremath{\mathrm{\,Ge\kern -0.1em V}}\xspace}
\newcommand{\mev}{\ensuremath{\mathrm{\,Me\kern -0.1em V}}\xspace}
\newcommand{\kev}{\ensuremath{\mathrm{\,ke\kern -0.1em V}}\xspace}
\newcommand{\ev}{\ensuremath{\mathrm{\,e\kern -0.1em V}}\xspace}
\newcommand{\gevc}{\ensuremath{{\mathrm{\,Ge\kern -0.1em V\!/}c}}\xspace}
\newcommand{\mevc}{\ensuremath{{\mathrm{\,Me\kern -0.1em V\!/}c}}\xspace}
\newcommand{\gevcc}{\ensuremath{{\mathrm{\,Ge\kern -0.1em V\!/}c^2}}\xspace}
\newcommand{\gevgevcccc}{\ensuremath{{\mathrm{\,Ge\kern -0.1em V^2\!/}c^4}}\xspace}
\newcommand{\mevcc}{\ensuremath{{\mathrm{\,Me\kern -0.1em V\!/}c^2}}\xspace}

\def\to                 {\ensuremath{\rightarrow}\xspace}

\def\gsim{{~\raise.15em\hbox{$>$}\kern-.85em
          \lower.35em\hbox{$\sim$}~}\xspace}
\def\lsim{{~\raise.15em\hbox{$<$}\kern-.85em
          \lower.35em\hbox{$\sim$}~}\xspace}

\def\AT#1     {\ensuremath{A_T^{#1}}\xspace}           

\def\C#1      {\ensuremath{\mathcal{C}_{#1}}}                       
\def\Cp#1     {\ensuremath{\mathcal{C}_{#1}^{'}}}                    
\def\Ceff#1   {\ensuremath{\mathcal{C}_{#1}^{\mathrm{(eff)}}}}        
\def\Cpeff#1  {\ensuremath{\mathcal{C}_{#1}^{'\mathrm{(eff)}}}}       
\def\Ope#1    {\ensuremath{\mathcal{O}_{#1}}}                       
\def\Opep#1   {\ensuremath{\mathcal{O}_{#1}^{'}}}                    

\begin{document}



\begin{titlepage}
\pagenumbering{roman}

\vspace*{-1.5cm}
\centerline{\large EUROPEAN ORGANIZATION FOR NUCLEAR RESEARCH (CERN)}
\vspace*{1.5cm}
\hspace*{-0.5cm}
\begin{tabular*}{\linewidth}{lc@{\extracolsep{\fill}}r}
\ifthenelse{\boolean{pdflatex}}
{\vspace*{-2.7cm}\mbox{\!\!\!\includegraphics[width=.14\textwidth]{figs/lhcb-logo.pdf}} & &}%
{\vspace*{-1.2cm}\mbox{\!\!\!\includegraphics[width=.12\textwidth]{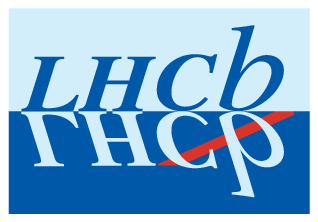}} & &}%
\\
 & & CERN-LHCb-DP-2012-003 \\  
 & & \today \\ 
 & & \\
\end{tabular*}

\vspace*{4.0cm}

{\bf\boldmath\huge
\begin{center}
Performance of the LHCb RICH detector at the LHC
\end{center}
}

\vspace*{2.0cm}

\begin{center}
The LHCb RICH group\footnote{Authors are listed on the following pages.}
\end{center}

\vspace{\fill}

\begin{abstract}
  \noindent
The LHCb experiment has been taking data at the Large Hadron
Collider (LHC) at CERN since the end of 2009.
  One of its key detector components is the Ring-Imaging
Cherenkov (RICH) system. This provides charged particle identification over a
wide momentum range, from 2--100 GeV$/c$.  
The operation and control, software, and online monitoring
of the RICH system are described. 
The particle identification performance is
presented, as measured using data from the LHC. 
Excellent separation of hadronic
particle types $(\pi,$\ K,\, p) is achieved.  
\end{abstract}

\vspace*{2.0cm}
\vspace{\fill}
\centerline {(To be submitted to EPJC)}
\vspace*{1.0cm}

\end{titlepage}


\newpage
%
%




\pagestyle{plain} 
\setcounter{page}{1}
\pagenumbering{arabic}


%



\centerline{\large\bf LHCb RICH collaboration}
\begin{flushleft}
\small
M.~Adinolfi$^{4}$, 
G.~Aglieri~Rinella$^{3}$, 
E.~Albrecht$^{3}$, 
T.~Bellunato$^{2}$, 
S.~Benson$^{7}$, 
T.~Blake$^{3,9}$, 
C.~Blanks$^{9}$, 
S.~Brisbane$^{10}$, 
N.H.~Brook$^{4}$, 
M.~Calvi$^{2,b}$, 
B.~Cameron$^{9}$, 
R.~Cardinale$^{1,a}$, 
L.~Carson$^{9}$, 
A.~Contu$^{10,e}$, 
M.~Coombes$^{4}$, 
C.~D'Ambrosio$^{3}$, 
S.~Easo$^{6,3}$, 
U.~Egede$^{9}$, 
S.~Eisenhardt$^{7}$, 
E.~Fanchini$^{2,b}$, 
C.~Fitzpatrick$^{7}$, 
F.~Fontanelli$^{1,a}$, 
R.~Forty$^{3}$, 
C.~Frei$^{3}$, 
P.~Gandini$^{10}$, 
R.~Gao$^{10}$, 
J.~Garra Tico$^{5}$,
A.~Giachero$^{2}$, 
V.~Gibson$^{5}$,
C.~Gotti$^{2}$, 
S.~Gregson$^{5}$,
T.~Gys$^{3}$, 
S.C.~Haines$^{5}$,
T.~Hampson$^{4}$, 
N.~Harnew$^{10}$, 
D.~Hill$^{10}$, 
P.~Hunt$^{10}$, 
M.~John$^{10}$, 
C.R.~Jones$^{5}$,
D.~Johnson$^{10}$, 
N.~Kanaya$^{3}$, 
S.~Katvars$^{5}$,
U.~Kerzel$^{5}$,
Y.M.~Kim$^{7}$, 
S.~Koblitz$^{3}$, 
M.~Kucharczyk$^{2,b}$, 
D.~Lambert$^{7}$, 
R.W.~Lambert$^{7,f}$, 
A.~Main$^{7}$, 
M.~Maino$^{2}$, 
S.~Malde$^{10}$, 
N.~Mangiafave$^{5}$,
C.~Matteuzzi$^{2}$, 
G.~Mini'$^{1}$, 
A.~Mollen$^{3}$, 
J.~Morant$^{3}$, 
R.~Mountain$^{11}$, 
J.V.~Morris$^{6}$, 
F.~Muheim$^{7}$, 
R.~Muresan$^{10,d}$, 
J.~Nardulli$^{6}$, 
P.~Owen$^{9}$, 
A.~Papanestis$^{6}$, 
M.~Patel$^{9}$, 
G.N.~Patrick$^{6}$, 
D.L.~Perego$^{2,b}$, 
G.~Pessina$^{2}$, 
A.~Petrolini$^{1,a}$, 
D.~Piedigrossi$^{3}$, 
R.~Plackett$^{9}$, 
S.~Playfer$^{7}$, 
A.~Powell$^{10}$, 
J.H.~Rademacker$^{4}$, 
S.~Ricciardi$^{6}$, 
G.J.~Rogers$^{5}$,
P.~Sail$^{8}$, 
M.~Sannino$^{1,a}$, 
T.~Savidge$^{9}$, 
I.~Sepp$^{9}$, 
S.~Sigurdsson$^{5}$,
F.J.P.~Soler$^{8}$, 
A.~Solomin$^{4}$, 
F.~Soomro$^{9}$, 
A.~Sparkes$^{7}$, 
P.~Spradlin$^{8}$, 
B.~Storaci$^{3,c}$, 
C.~Thomas$^{10}$, 
S.~Topp-Joergensen$^{10}$, 
N.~Torr$^{10}$, 
O.~Ullaland$^{3,b}$, 
K.~Vervink$^{3}$, 
D.~Voong$^{4}$, 
D.~Websdale$^{9}$, 
G.~Wilkinson$^{10}$, 
S.A.~Wotton$^{5}$,
K.~Wyllie$^{3}$, 
F.~Xing$^{10}$, 
R.~Young$^{7}$.\bigskip

{\footnotesize \it
$ ^{1}$Sezione INFN di Genova, Genova, Italy\\
$ ^{2}$Sezione INFN di Milano Bicocca, Milano, Italy\\
$ ^{3}$European Organization for Nuclear Research (CERN), Geneva, Switzerland\\
$ ^{4}$H.H. Wills Physics Laboratory, University of Bristol, Bristol, United Kingdom\\
$ ^{5}$Cavendish Laboratory, University of Cambridge, Cambridge, United Kingdom\\
$ ^{6}$STFC Rutherford Appleton Laboratory, Didcot, United Kingdom\\
$ ^{7}$School of Physics and Astronomy, University of Edinburgh, Edinburgh, United Kingdom\\
$ ^{8}$School of Physics and Astronomy, University of Glasgow, Glasgow, United Kingdom\\
$ ^{9}$Imperial College London, London, United Kingdom\\
$ ^{10}$Department of Physics, University of Oxford, Oxford, United Kingdom\\
$ ^{11}$Syracuse University, Syracuse, NY, United states of America\\
\bigskip
$ ^{a}$Universit\`{a} di Genova, Genova, Italy\\
$ ^{b}$Universit\`{a} di Milano Bicocca, Milano, Italy\\
$ ^{c}$Now at Physik-Institut, Universit\"at Z\"urich, Z\"urich, Switzerland\\
$ ^{d}$Now at Horia Hulubei National Institute of Physics and Nuclear Engineering, Bucharest-Magurele, Romania\\
$ ^{e}$Now at Sezione INFN di Cagliari, Cagliari, Italy\\
$ ^{f}$Now at Nikhef National Institute for Subatomic Physics and VU 
University Amsterdam, Amsterdam, The Netherlands \\
}
\end{flushleft}

\section{Introduction}
\label{sec:Introduction}

\begin{figure}[t]
  \centering 
\includegraphics*[width=160mm]{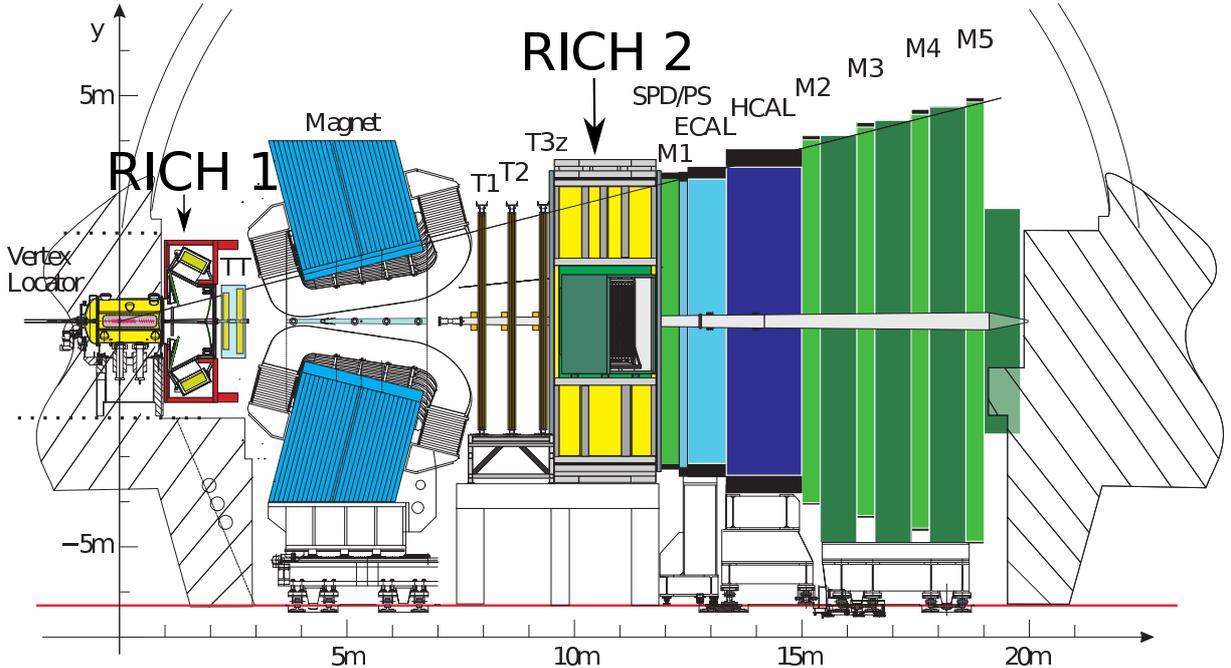}
  \caption{\small Side view of the LHCb spectrometer, with
  the two RICH detectors indicated}
  \label{fig:spect}
\end{figure}

LHCb~\cite{Detector}
is one of the four major experiments at the LHC, and is dedicated to the study of
CP violation and the rare decay of heavy flavours.  
It is a forward spectrometer designed to accept forward-going 
$b$-and $c$-hadrons produced in proton-proton collisions. 
The layout of the spectrometer is shown in Fig.~\ref{fig:spect}.
The subdetectors of LHCb are described in detail in Ref.\cite{Detector}.

The RICH system of the LHCb experiment provides charged
particle identification over a wide momentum range, from 2 to 100\,GeV$/c$.  
It consists of two RICH detectors that cover
between them the angular acceptance of the experiment, 15--300\,mrad with
respect to the beam axis.  
The LHC accelerator started at the end of 2009 and ran at a centre-of-mass 
energy of 7\,TeV until the end of 2011, followed by 8\,TeV in 2012.  
The luminosity rapidly increased and at the end of 2010  reached the 
nominal operating value for the LHCb experiment, 
$2\times 10^{32}$\,cm$^{-2}$\,s$^{-1}$.
  This paper describes the performance of the RICH system and also its
 alignment and calibration using data.  
Many LHCb results have already fully exploited the RICH 
capabilities~\cite{results}.  
  
The paper is structured as follows:  the requirements for particle
identification are discussed in Sect.~1, and a brief
description of the RICH detectors is given in Sect.~2.  The alignment and
calibration of the detectors are described in Sect.~3.
 Section 4 gives an overview of the software used in the RICH reconstruction, 
particle identification and detector performance,
followed in Sect.~5 by the conclusions.

\subsection{Requirements for particle identification}

The primary role of the RICH system is the identification of charged 
hadrons ($\pi$, K, p).  

One of the major requirements for charged hadron identification in a
flavour-physics experiment is for the reduction of combinatorial
background. Many of the interesting decay modes of $b$- and $c$- flavoured
hadrons involve hadronic multibody final states.  At hadron colliders like the
LHC, the most abundant produced charged particle is the pion.
The heavy flavour decays of interest typically contain a number of kaons, 
pions and protons.
It is therefore important in reconstructing the invariant mass of the decaying 
particle to be able to select the charged hadrons of interest in order to 
reduce the combinatorial background.  

\begin{figure}[tp]
\begin{center}
\includegraphics[scale=0.4]{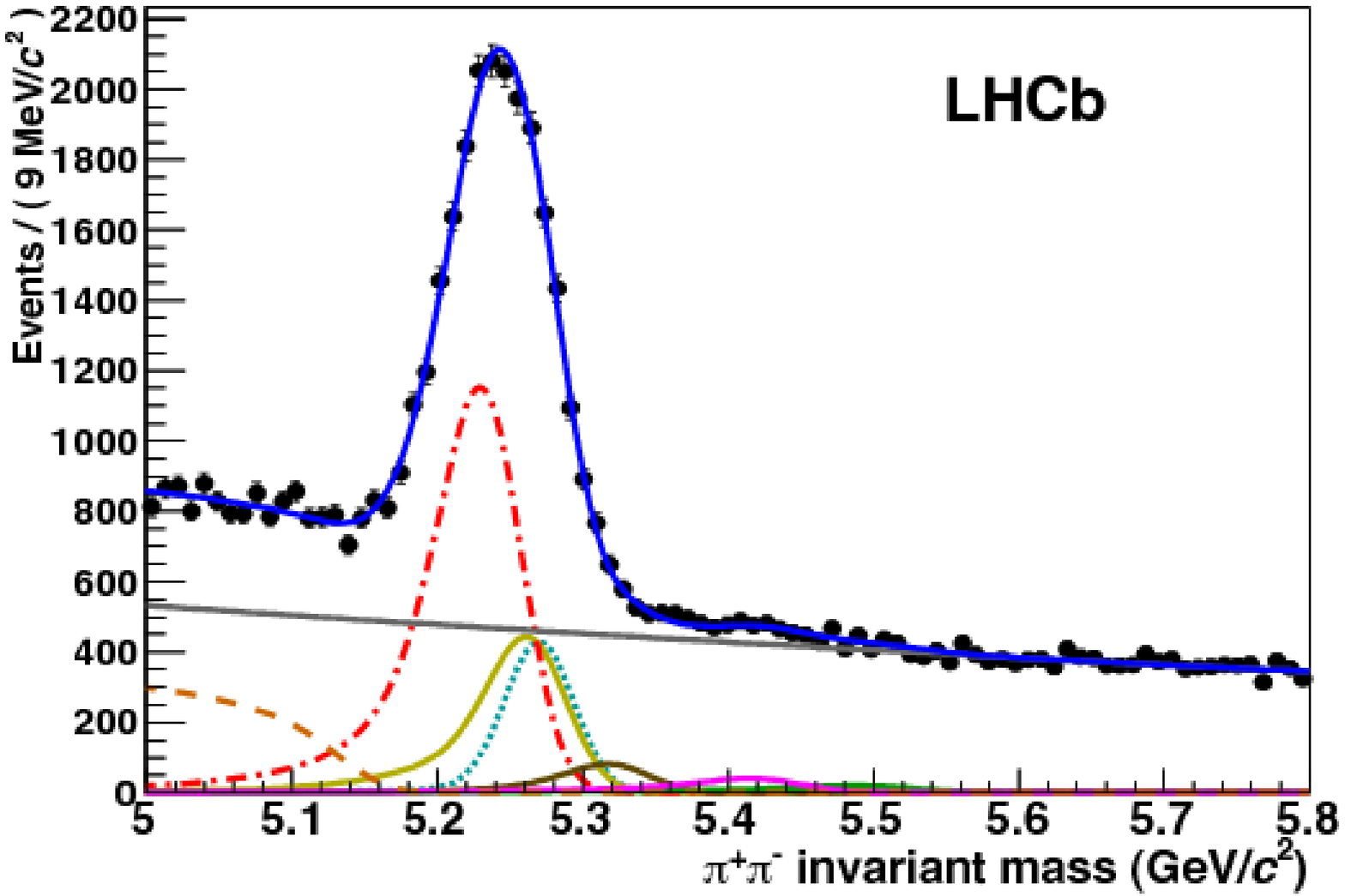}
\includegraphics[scale=0.4]{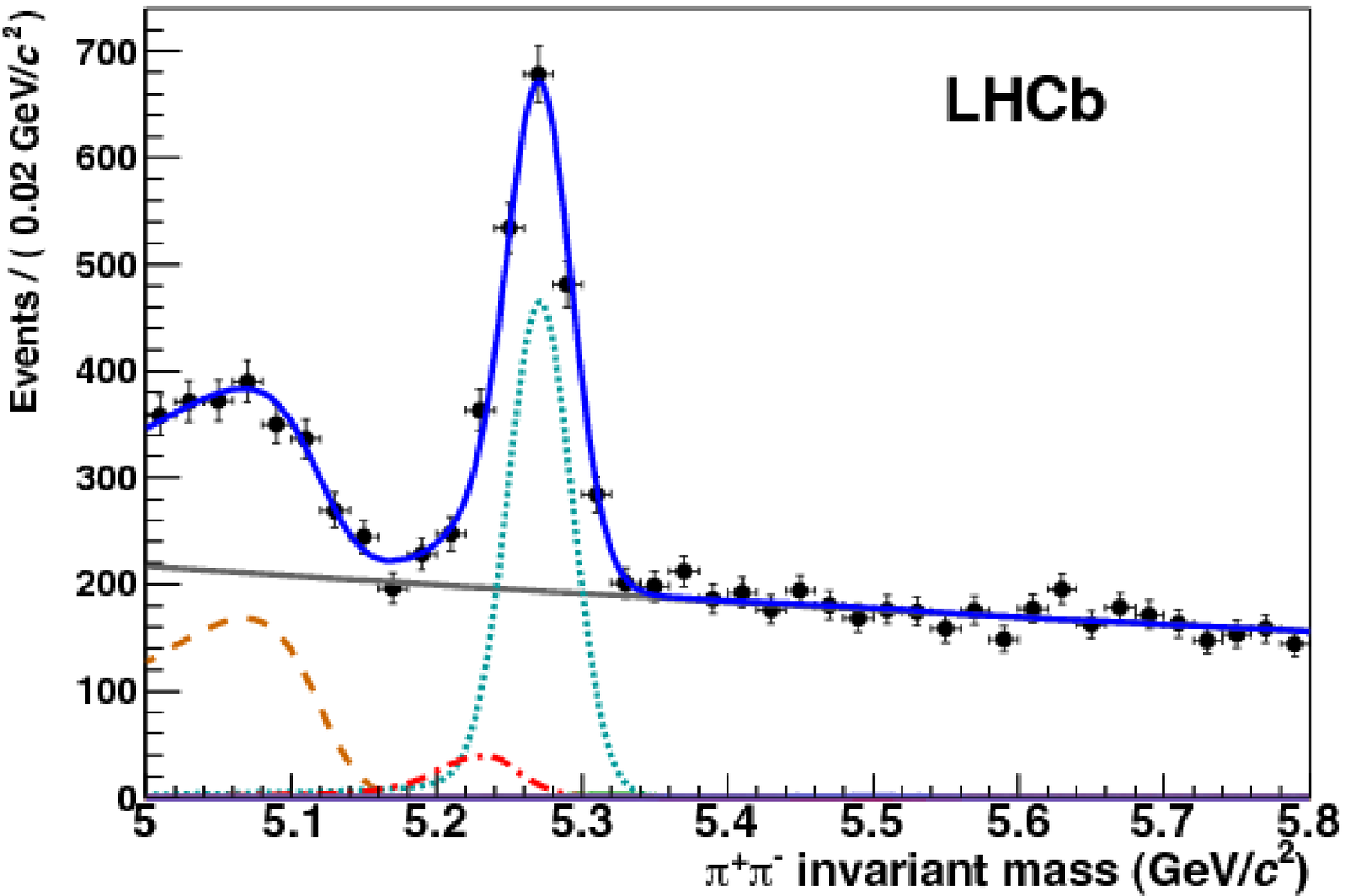}
\caption{ Invariant mass distribution for B$\rightarrow h^+h^-$ 
decays~\cite{vagnoni}  
  in the LHCb data before the use of the RICH information (left), 
and after applying RICH particle identification (right). 
The signal under study is the decay 
B$^0\rightarrow\pi^+\pi^-$, represented by the turquoise dotted line.
The contributions from different $b$-hadron decay modes 
(B$^0 \rightarrow$ K$\pi$ red dashed-dotted line,
B$^0\rightarrow $3-body orange dashed-dashed line,
B$_s \rightarrow$ KK yellow line,
B$_s \rightarrow$ K$\pi$ brown line,
$\Lambda_b \rightarrow$ pK purple line,
$\Lambda_b \rightarrow$ p$\pi$ green line),
 are eliminated by positive identification of
pions, kaons and protons and only the signal and two background contributions 
remain visible in the plot on the right.
The grey solid line is the combinatorial background}
  \label{fig:bhh-data}
\end{center}
\end{figure}

The second major use of the particle identification information is to 
distinguish final states of otherwise identical topology.  
An example is the two-body hadronic
decays, B$\rightarrow h^+h^-$, where $h$ indicates a charged 
hadron~\cite{vagnoni}.  
In this case there are many contributions, as
illustrated in Fig.~\ref{fig:bhh-data}, including 
B$^0\rightarrow\pi^+\pi^-$,
B$^0_s\rightarrow$ K$^+$K$^-$, and other decay modes of the B$^0$, 
B$^0_s$ and
$\Lambda_b$.
A signal extracted using only kinematic and vertex-related cuts is a sum over 
all of the decay modes of this type 
(Fig.~\ref{fig:bhh-data} left),
 each of which will generally have a different CP asymmetry.
For a precise study of CP-violating effects, it is crucial to
separate the various components. This is achieved by exploiting the 
high efficiency of the RICH particle identification
 (Fig.~\ref{fig:bhh-data} right).

Another application of charged hadron identification is for an efficient 
flavour tagging~\cite{tagging}.  
When studying CP asymmetries or particle-antiparticle oscillations, 
knowledge of the production 
state of the heavy-flavoured particles is required.
This can be achieved by tagging the
particle/antiparticle state of the accompanying hadron.
Heavy-flavoured particles are predominantly produced in pairs. 
One of the most powerful means of tagging the production state is by
identifying charged kaons produced in the $b\rightarrow c\rightarrow s$ cascade
decay of the associated particle. 
Such tagged kaons (as well as kaons from the $b$ fragmentation when a 
B$^0_s$  is created), have a soft momentum
distribution, with a mean of about 10\,GeV/$c$.
Particle identification down
to a few GeV/$c$ can therefore significantly increase the tagging power of 
the experiment.

The typical momentum of the decay products in two-body $b$ decays is about
50\,GeV/$c$. The requirement of maintaining a high efficiency for the
reconstruction of these decays leads to the need
for particle identification up to at least 100\,GeV/$c$.  
The lower momentum limit of
about 2\,GeV/$c$ follows from the need to identify decay products from 
high multiplicity B decays and also from the fact that particles below this 
momentum will not pass through the dipole magnetic field (4 Tm) of the LHCb 
spectrometer.  

A further example of the requirements for particle identification in LHCb is 
its use in the trigger.  
LHCb has a high performance trigger system ~\cite{trigger}, that reduces the 
event rate from the 40\,MHz bunch crossing frequency down to about 2\,kHz 
that can be written to storage.  
This is achieved in two steps. 
The first trigger level is implemented in hardware and is
based on high transverse energy deposits in the calorimeter and high 
transverse momentum detected by the muon system,
to reduce the rate to 1\,MHz. 
All detectors are then read out into a
CPU farm where a high level trigger (HLT, see 
 Fig.~\ref{fig:MonDataFlow} )
decision is made fully in software.
The RICH reconstruction is fast enough to contribute to this 
trigger. An example is the online selection of the $\phi$ particle, 
which is present in many of the decay modes of interest.

\section{The RICH detectors}
\label{sec:Detector}


\subsection{Detector description }
\label{sec:_RICHDetector}

A description of the LHCb RICH detectors is given in Ref.\cite{Detector}.  
Only the major features are summarized here.  
In the forward region, covered by the LHCb spectrometer, there is a strong 
correlation between
momentum and polar angle, with the high-momentum particles produced
predominantly at low polar angles.
As a result, the RICH system has two detectors.  
RICH\,1  covers the low and intermediate momentum region 2 - 40\,GeV$/c$
over the full spectrometer angular acceptance of 25--300\,mrad. 
The acceptance is limited
at low angle by the size of the beampipe upstream of the magnet.
RICH\,2 covers the high-momentum region 15--100\,GeV$/c$, over the angular 
range 15--120\,mrad.  

To limit its overall volume, RICH\,1 is placed as close as possible to the 
interaction region.
It is located immediately downstream of the silicon-microstrip vertex locator 
(VELO), as shown in Fig.~\ref{fig:spect}.  
To minimize the material budget there is no separate entrance window, 
and the RICH\,1 gas enclosure is sealed directly
to the exit window of the VELO vacuum tank.  
The downstream exit window is constructed from a low-mass
carbon-fibre/foam sandwich. 
RICH\,2 is placed downstream of the
magnet, since the high momentum tracks it measures are less affected by the
magnetic field. In this way it can be placed after the downstream
tracking system in order to reduce material for the measurement of the
charged tracks. 
The entrance and exit windows are again a foam sandwich construction and
 skinned with carbon-fibre and aluminium, respectively.

Both RICH detectors have a similar optical system, with a tilted spherical
focusing primary mirror, and a secondary flat mirror to limit the length of the
detectors along the beam direction.  Each optical system is divided
into two halves on either side of the beam pipe, with RICH\,1 being divided
vertically and RICH\,2 horizontally.
The vertical division of RICH\,1 was necessitated by the requirements of
magnetic shielding for the photon detectors, due to their close proximity to the
magnet.  
The spherical mirrors of RICH\,1 (4 segments) are constructed in four 
quadrants, with
carbon-fibre structure, while those of RICH\,2 (56 segments), and all flat 
mirrors (16 and 40 segments in RICH\,1 and RICH\,2 respectively), are
tiled from smaller mirror elements, employing a thin glass substrate.  A
reflectivity of about 90\% was achieved for the mirrors, averaged over the
wavelength region of interest, 200--600\,nm.  The total material budget for
RICH\,1 is only about 8\% $X_0$ within the experimental
acceptance, whilst that of RICH\,2 is about 15\% $X_0$.

Fluorocarbon gases at room temperature and pressure are used as
Cherenkov radiators;
C$_4$F$_{10}$ in RICH\,1 and CF$_4$ in RICH\,2  
 were chosen for their low dispersion. 
The refractive index is respectively  1.0014 and 1.0005 at 0$^{\rm o}$~C, 101.325~kPa
 and 400 nm.
About 5\% CO$_2$ has been added to the CF$_4$ in order to quench scintillation 
in this gas~\cite{scintillation}.

The momentum threshold for kaons to produce Cherenkov light in C$_4$F$_{10}$ is
9.3\,GeV$/c$. Particles below this momentum would only be identified as
kaons rather than pions in veto mode, i.e. by the lack of Cherenkov light
associated to the particle.  To maintain positive identification at low momentum
and in order to separate kaons from protons, a
second radiator is included in RICH\,1: a 50 mm thick wall made of 16 tiles of 
silica aerogel~\cite{Aerogel} at the entrance to RICH\,1.
The refractive index is $n=1.03$ and the light scattering length is around 
50 mm at 400 nm in pure N$_2$.
The aerogel is placed in the C$_4$F$_{10}$ gas volume and a thin glass filter 
is used on the downstream face to limit the chromatic dispersion.

The Cherenkov photons emitted by charged particles passing through the RICH 
radiators are
focused into ring images on the photon detector planes, situated outside of the
spectrometer acceptance.  A novel hybrid photon detector (HPD) was developed in
collaboration with industry specifically for application in the LHCb RICH
system~\cite{HPD}.
The HPDs employ vacuum tubes with a 75\,mm active diameter, with a
quartz window and multialkali photocathode.  The photoelectrons are focused
onto a silicon pixel array, using an accelerating voltage of -16\,kV.
The pixel array is arranged in 32 columns and 32 rows, giving a total of 1024
pixels per tube.  
The pixel size is 2.5$\times$2.5 mm$^2$ at the level of the photocathode.
A total of 484 HPDs are close-packed to cover the four
photodetector planes.
 Two planes are employed in each RICH, with 196 tubes used in RICH\,1 and 288
in RICH\,2.  
The photodetector planes are separated from the radiator gas
volumes by quartz windows, and the photodetector volumes are maintained in 
an atmosphere of CO$_2$.  
The front-end electronics chip is encapsulated within the HPD vacuum tube, and
bump-bonded to the silicon pixel sensor, which results in extremely low noise
(typically 150~e$^-$ RMS per pixel for a signal of 5000~e$^-$~\cite{lownoise}).
  The tubes also feature high detection efficiency, with an active area 
fraction of about 82\%.
The quantum efficiency is about 30\% at 270\,nm.

\subsection{Detector operation }
\label{sec:_RICHoperation}

The operation of the RICH detectors is fully automated and is controlled by the
Experiment Control System (ECS)~\cite{RichDCS}.
The RICH ECS has been built using components from the Joint Controls Project
framework~\cite{JCOP}, developed by CERN and the four main LHC 
experiments. 
The ECS uses predefined sequences for normal detector operation, allowing
non-experts to operate the detectors. Automated actions protect the
equipment when monitored parameters fall outside the range of accepted values. 
Sensors are used as input to the LHCb Detector Safety System which
put the detectors in a safe state in case of a major malfunction of the control
system.


The RICH ECS also collects environmental information that is required by the
RICH reconstruction software. There are systems to monitor movements of the 
RICH mirrors, monitor the quality of the gas radiators, and log the temperature
and pressure of the radiators in order to correct the refractive index. 
Changes in temperature and pressure, which necessitate the re-calculation of 
the refractive index of the gas radiators,
are automatically propagated to the Conditions 
Database~\cite{CondDB}. 

\begin{figure}
\begin{center}
\includegraphics[scale=0.5]{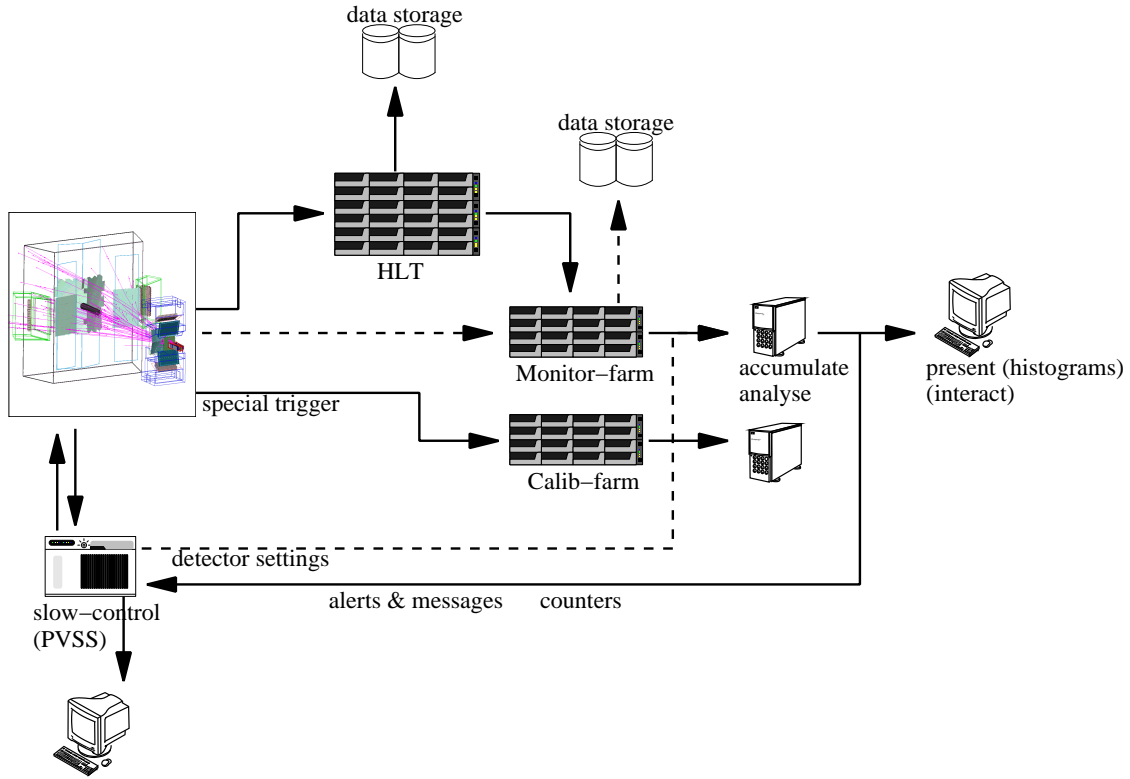}
\caption{RICH data-flow through the online system. 
Events selected by the L0 trigger are sent to the High Level Trigger (HLT) farm
and, if they pass this trigger requirements, are sent to storage. 
A fraction of these events (typically 10\%) is also sent to the 
monitoring farm. Online monitoring algorithms examine the data for 
irregularities and send messages to the slow-control (ECS) that can trigger 
automatic actions. Special triggers are sent directly to the calibration farm 
bypassing the High Level Trigger
  }
\label{fig:MonDataFlow}
\end{center}
\end{figure}

The RICH detectors and the data recorded are monitored
at several stages of the data-acquisition and reconstruction
chain to identify any potential problems
as early as possible.
Figure~\ref{fig:MonDataFlow} illustrates the online data-flow,
highlighting the dedicated monitoring and calibration farms for
analyses using fully reconstructed events.

Low-level processes monitor the data integrity during data recording 
by cross-checking the various data-banks and reporting any irregularities.
Higher-level monitoring algorithms
use a neural network~\cite{nnet}
to identify Cherenkov rings
using information from the RICH detectors only. 
On rare occasions, an individual HPD may 
lose synchronisation with the rest of the detector and transmit
spurious data for each event. 
It is found that the performance of the particle identification is affected 
only marginally by a few units of malfunctioning HPDs\footnote{The number of malfuctioning HPDs is considered acceptable if it is 
less than 6 peripheral tubes, or one central tube.},
and it is usually more 
effective to continue recording data and reset those affected front-end 
components during the next run initialization. 
In order to prevent inefficiencies during data-taking due to anomalously
busy events, the online monitoring task automatically detects these cases and 
the read-out electronics discards all data prior to transmission.

Special calibration triggers can be sent to the photodetectors 
during normal data-taking to activate a pre-defined test pattern of hits.
This provides a continuous test of the response of all photodetectors,
especially in low-occupancy regions. 
As these calibration triggers are sent during 
gaps in the bunch-train structure of the LHC beam, these events 
contain no activity related to proton-proton interactions. 
These ``empty'' events can also be used to evaluate
noise that would be present in the detector during data-taking.

The online monitoring allows the full
event reconstruction of a sizeable fraction of the events being recorded
to be processed online.
This allows the monitoring of the reconstructed Cherenkov angle, as well 
as the alignment of the RICH detectors with respect to the tracking system.

\newcommand {\defuni}[1] {\ifmmode \mathrm{#1} \else $\mathrm{#1}$ \fi}
\newcommand {\um}[1]     {\defuni{\; #1}}

\def\cfourften{\ensuremath{\mathrm{C}_{4}\mathrm{F}_{10}}\xspace}

\section{Alignment and calibration}
\label{sec:Operations}

The tasks of spatial alignment of the RICH detectors and the calibration of the
refractive indices of the radiators are performed with data using
high momentum charged particles. 
In addition, the alignment of several
mirror segments and the purity of the gas radiators are also monitored
using systems that can provide information independently and during
periods when there are no collisions.   


\subsection{Time alignment}
In order to maximise the photon collection efficiency of the RICH, 
the HPD readout must be synchronised with the LHC bunch crossing to within a 
few nanoseconds. 
This procedure is referred to as "time alignment" in the following.
Individual HPDs vary in timing due to variations in
drift time of the electrons within the silicon sensor. 
HPD readout is triggered by a 25 ns wide strobe pulse distributed by the LHCb 
network of optic fibres and detected by the RICH Level-0 (L0) 
boards~\cite{adinolfi}. 
A RICH L0 board supervises the triggering, timing and control of the HPDs, 
with two HPDs serviced by a single board. 
HPDs that share a L0 board were chosen to have similar timing
characteristics in order to optimise the time alignment.

Three features on the timing profile are defined: the $turn$-$on$ 
point is the delay between optimal readout efficiency and the strobe
pulse at which 90\% of the peak photon collection efficiency is observed, 
the $turn$-{\it off} point is the delay at which the profile drops below 90\%, 
and the midpoint is the average delay between these values.

\begin{figure}
\centering
\includegraphics[scale=0.4]{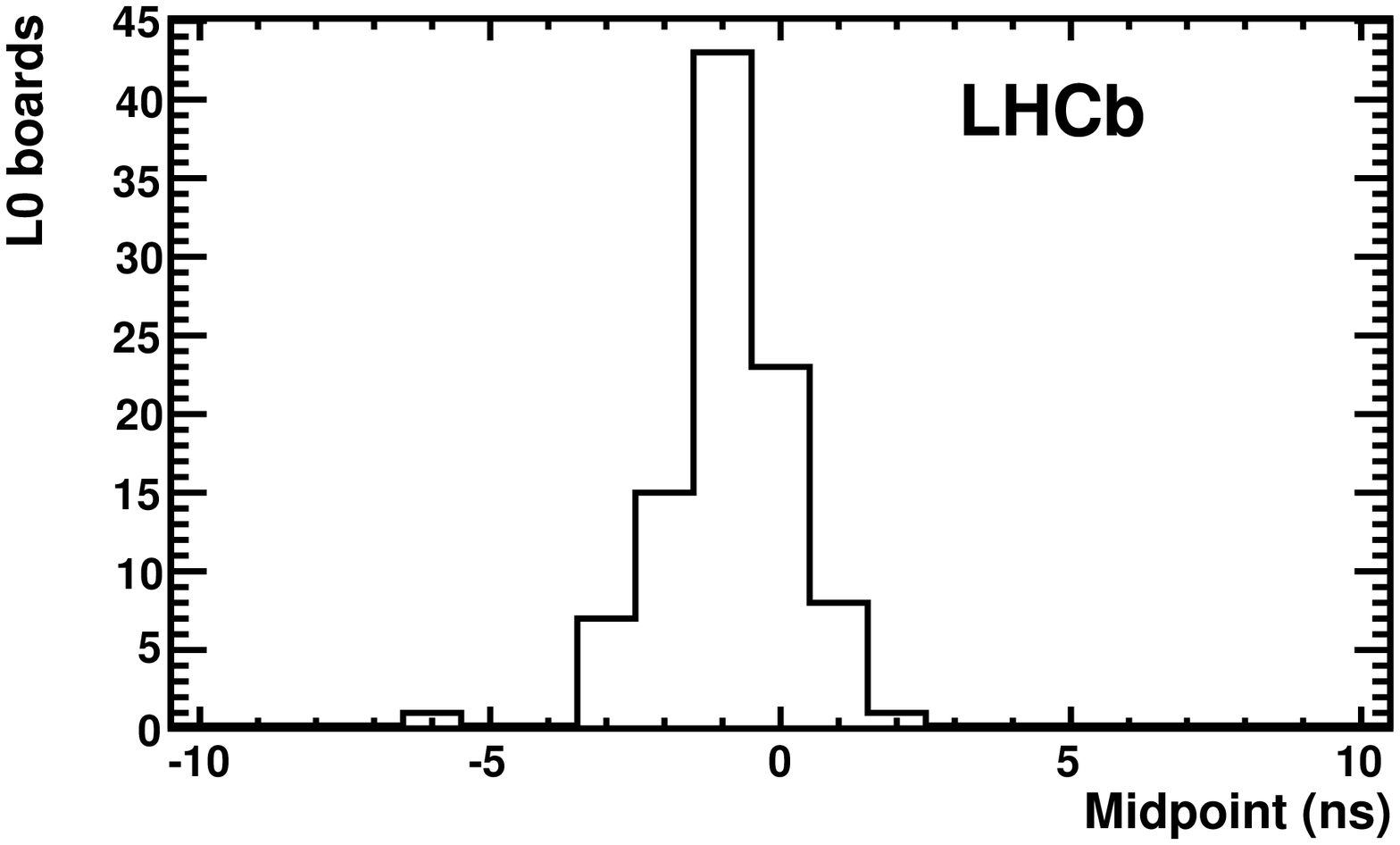}
\includegraphics[scale=0.4]{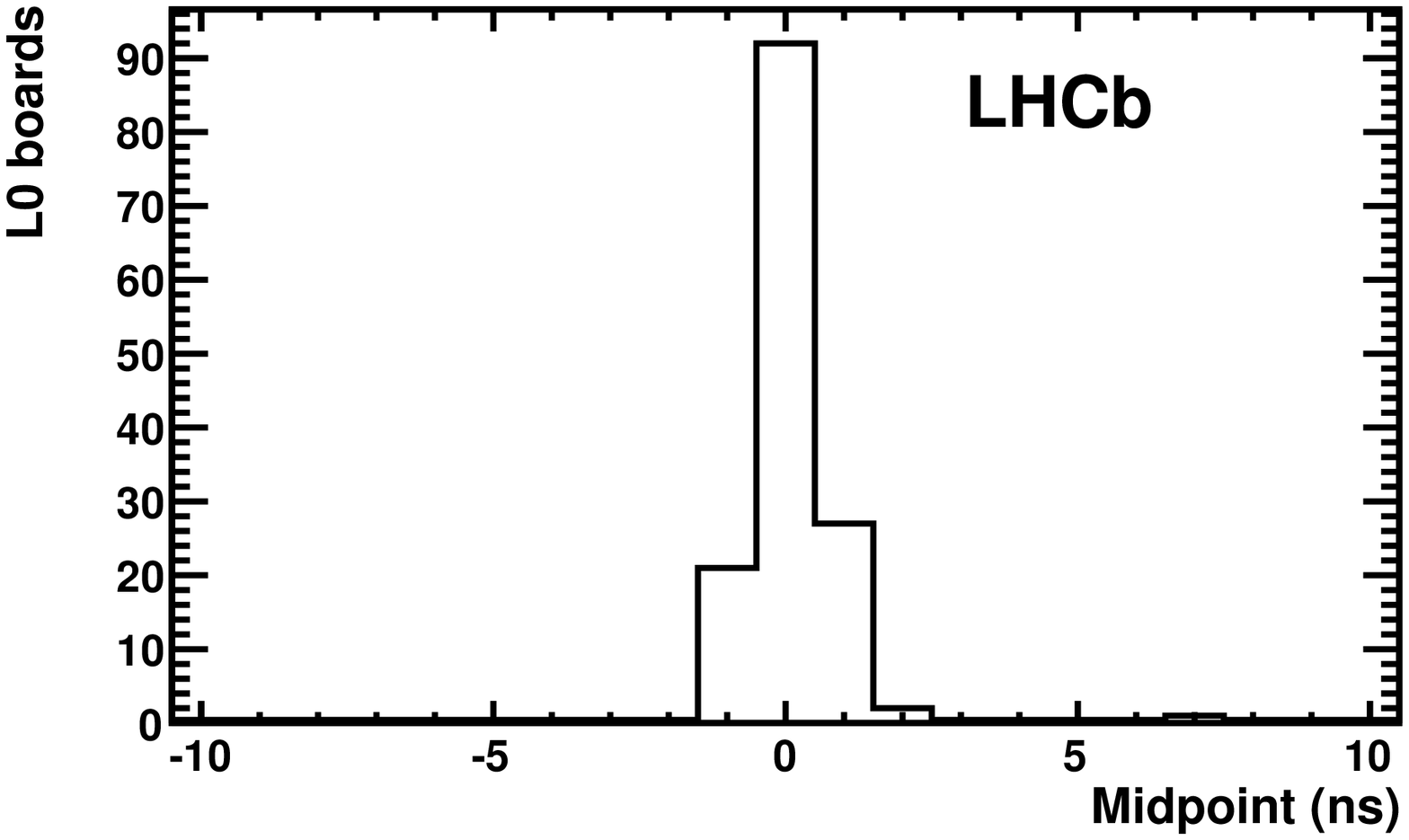}
\caption[Time alignment scans]{Distribution of the midpoints in RICH\,1
(left) and RICH\,2 (right) after time alignment with pp collisions. 
The RMS deviations of the HPDs are approximately 1 ns}
\label{timing}
\end{figure}

The global time alignment of L0 boards has been performed in several
 steps both
prior to and during running at the LHC. The initial alignment was performed
in the absence of beam using a pulsed laser, resulting in a relative alignment of
the HPDs in each photodetector plane. The global time alignment to 
the LHCb experiment is achieved with pp collisions using the LHCb first level
 trigger.
The distribution of the midpoints can be seen in Fig.~\ref{timing},
showing that the HPDs have been time aligned to $\sim$ 1~ns.

\subsection{Magnetic distortions}
\label{sec:MDMS}
Inside an HPD, photoelectrons travel up to 14 cm 
from the photocathode to the silicon anode. 
This device is therefore sensitive to stray 
magnetic fields from the LHCb spectrometer magnet.
All HPDs in RICH\,1 and RICH\,2 experience some residual 
fringe field:
the magnetic shields surrounding the HPDs reduce it from initial values of
 up to 60(15) mT in RICH\,1(RICH\,2), to a maximum value of $\simeq$ 2.4 mT 
in RICH\,1, and values ranging between 0.2 - 0.6 mT in RICH\,2.
 The resulting distortion of the images are 
mapped and corrected when reconstructing Cherenkov angles.

A characterisation procedure has been developed to correct for magnetic 
distortion effects and restore the optimal resolution. 
Different implementations are used for RICH\,1 and RICH\,2 as 
the two detectors have different geometries and experience different
field configurations.

\subsubsection{RICH\,1}
The distortions of the HPD images are corrected using a dedicated calibration
system.
The mapping system consists of two identical hardware arrangements, one for 
each of the upper and lower HPD boxes \cite{MDMSyst}.  
Each system consists of an array of green light-emitting diodes mounted on a 
carbon-fibre support that spans the width of the HPD box.  This ``light bar"
attaches at each end to a pair of synchronised stepper-motors that 
facilitates the
illumination of the entire HPD array.  The light bar comprises 19 LED 
units each
with an array of $5\times28$ LEDs, 2.5~mm apart.  A passive collimator unit
is mounted flush to the LED array such that light from each LED is 
channeled down an individual collimator with 0.3 mm aperture.


\begin{figure}[htbp]
\centering
\includegraphics[width=0.45 \textwidth]{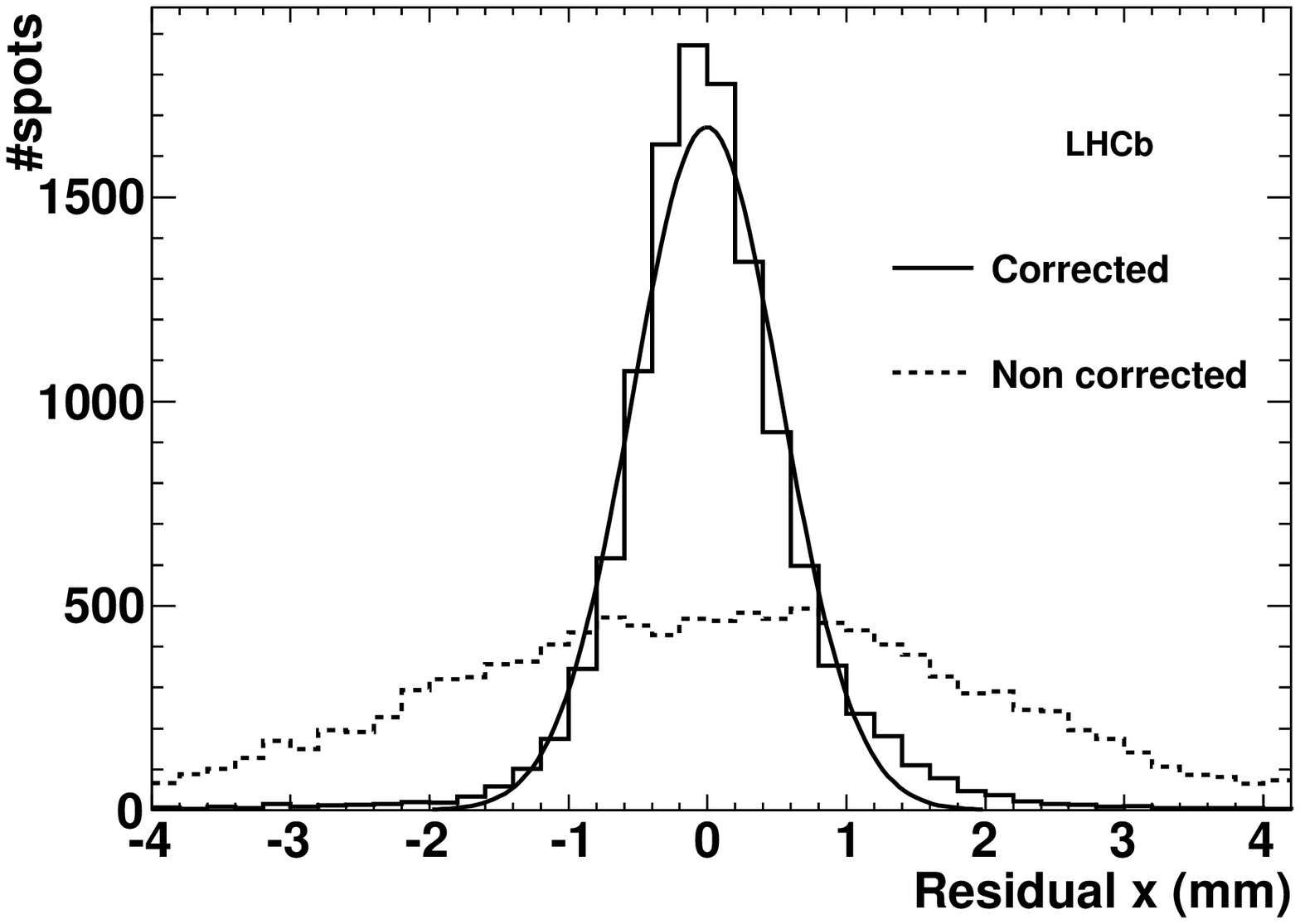}
\includegraphics[width=0.45 \textwidth]{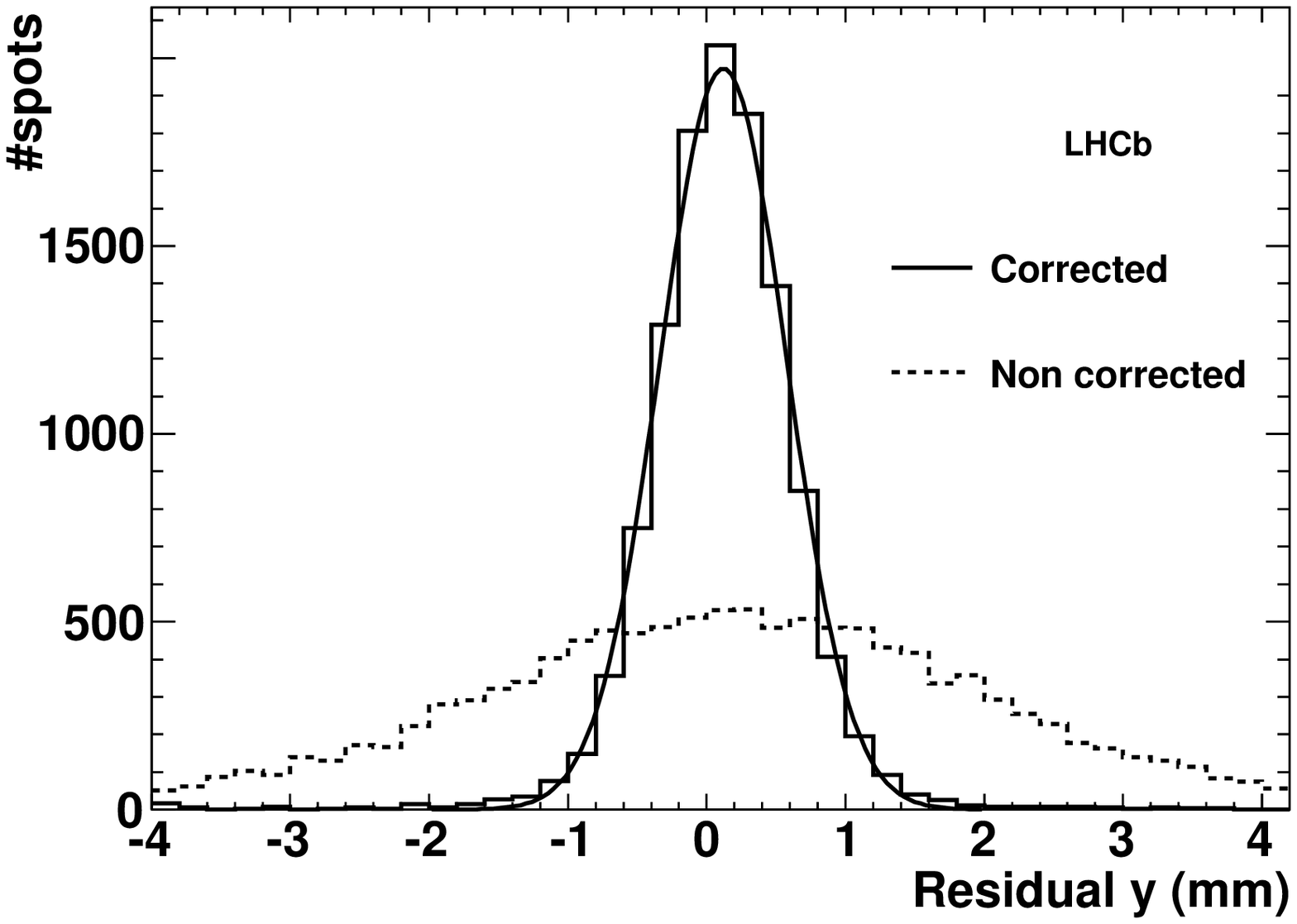}
\caption{
Spatial residuals demonstrating the resolution with which the light 
spots of the test pattern in RICH\,1 are identified.  The plot
shows the distance from the measured light spot centre
to the nearest test point. 
The dotted and solid lines are before and after the calibration respectively,
 along the $x$ direction (left) and along $y$ (right) of the anode plane,
 projected on the photocathode plane.
The solid line is the Gaussian fit
}
\label{fig:mdmSpaceRes}
\end{figure}

The distortion is mapped by comparing the pattern of
light spots in magnet-on and magnet-off data.  The direction of the magnetic
field is predominantly longitudinal with respect to the tube axis.  
The field effect causes
a rotation of the image about the central axis of the HPD, and 
 a modest expansion of the image.  
The residual transverse
component of the field displaces the centre of the photocathode image. 

The result of the parametrisation is demonstrated in 
Fig.~\ref{fig:mdmSpaceRes}, showing  
the residual positional uncertainty due to the magnetic distortion 
after the correction procedure.
The resolution after correction is $\sigma\simeq 0.2$~pixel (0.5~mm), 
significantly smaller than the
irreducible uncertainty of $\sigma \simeq 0.29\um{pixel}$ (0.72~mm),
originating from the finite HPD pixel size.

\subsubsection{RICH\,2} 
The magnetic distortion correction for RICH\,2 uses a system based on the
projection of a light pattern onto the plane of HPDs using a commercial light
projector.  The projected pattern 
is a suitable grid of light spots. 
The algorithm to reconstruct the position of a light spot
builds a cluster of hits and the cluster centre is calculated. 
A resolution better than the pixel size is achieved. 
Comparing the position of
the light spots with and without the magnetic field makes it possible to
measure and parametrise the magnetic field 
distortions~\cite{bi:Thesis,bi:TGnote}.

\begin{figure}[htbp]
\centering
\includegraphics[scale=0.40]{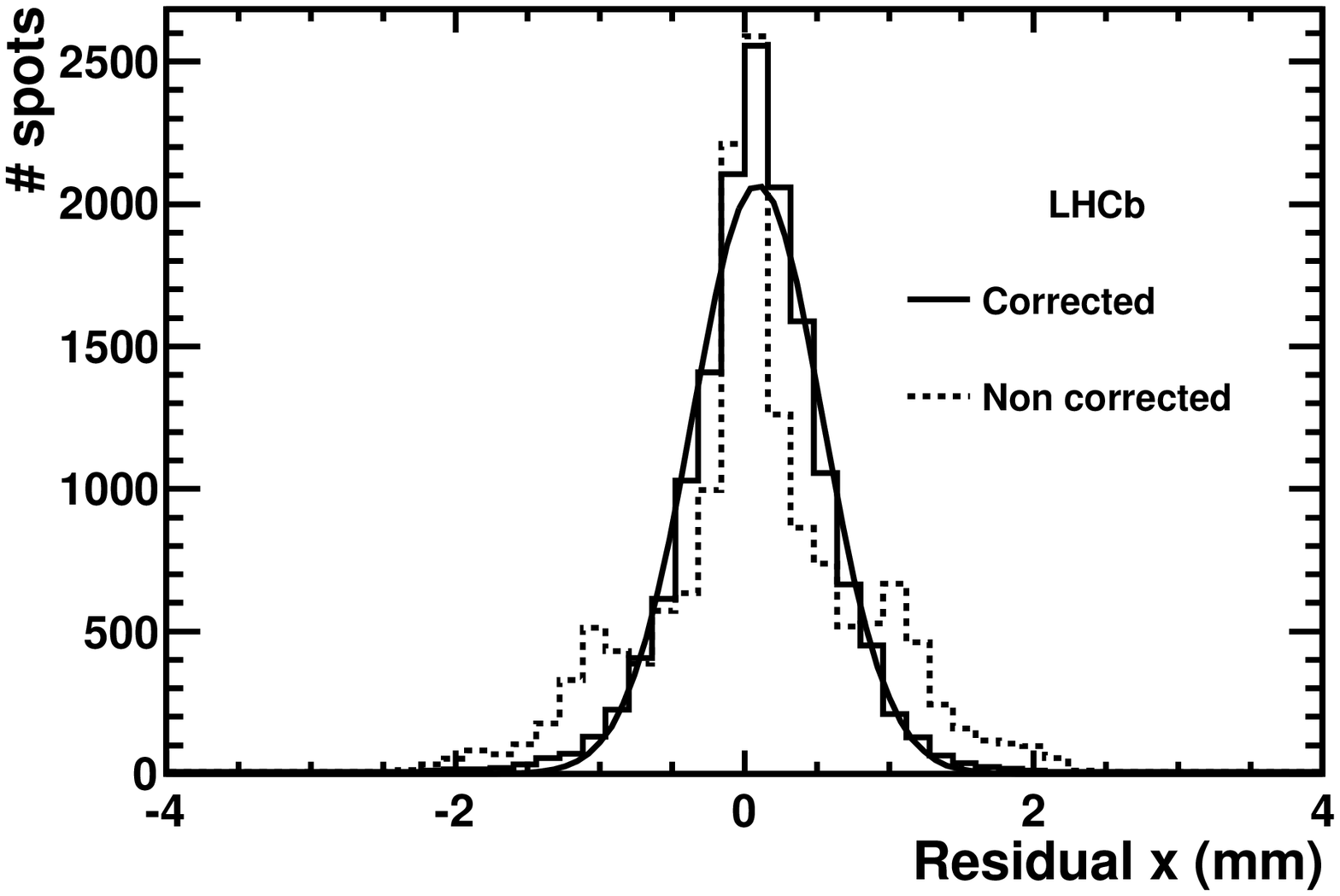}
\includegraphics[scale=0.40]{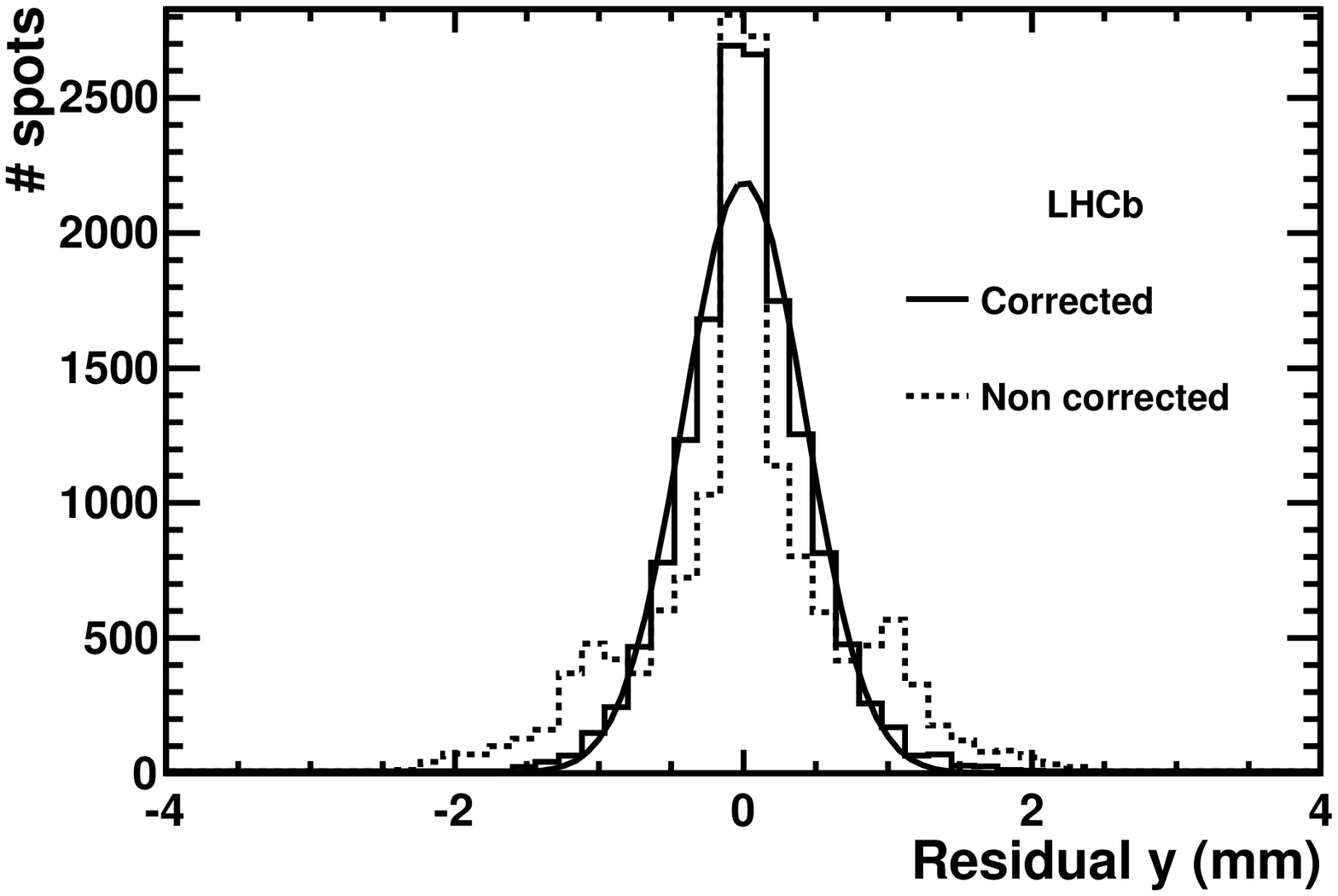}
\caption{
Spatial residuals demonstrating the resolution with which the light 
spots of the test pattern in RICH\,2 are identified.  
The plot shows the distance from the measured light spot centre
to the nearest test point. 
The dotted and solid lines are before and after the calibration respectively.
Most of the photodetectors of RICH\,2 are in a region 
free from magnetic field residual values (region around $x$=0 of the dotted 
histogram). 
Where these are different from zero, 
the distorsions induced are visible in the two 
satellite peaks of opposite sign (the magnetic field changes sign in 
the upper and lower part of the photodetector matrix plane).
The left plot is the measurement along the $x$, the right 
plot along $y$ of the anode plane, projected on the photocathode plane.
The solid line is the Gaussian fit
 }
\label{fig:cgrot}
\end{figure}

The distortion mainly consists of a small rotation (on average   
\mbox{$\lesssim 0.1\um{rad}$}) 
 of the test spots around the HPD axis. 
This rotation varies from HPD to HPD, depending on the HPD position.
No measurable variation of the radial coordinate of the light spots was detected. 
The parameters extracted using either orientation of the 
magnetic field also apply, with a sign inversion, to the opposite
magnetic field polarity.
By applying the correction procedure, the resolution on the position 
of the light spot improves from 
$\sigma\simeq 0.33 \um{pixel}$
to $\sigma\simeq 0.19 \um{pixel}$ (0.47~mm)
(see Fig.~\ref{fig:cgrot}).  
For comparison the pixel size resolution is $\sigma = 0.29\um{pixel}$ (0.72~mm).

\subsection{Detector alignment}
\label{sec:Software:mirrors}

In order to reconstruct the Cherenkov angle associated with the individual 
photons as accurately as possible, a number of components
must be aligned with an accuracy of 0.1~mrad with respect to the LHCb tracking 
system. 
The alignment procedure calculates
the misalignments of the various detector components in a sequential process. 
First, the whole RICH detector is aligned with the global LHCb 
coordinate system, followed by each detector half, each mirror segment 
and finally each HPD. 
This includes aligning the silicon sensors inside the HPDs to the whole 
RICH detector. 
One has to know the position of the centre of
each HPD photocathode on the anode.
The silicon sensors are aligned by mapping an image of the photocathode.
This procedure does not require the reconstruction of the Cherenkov angle. 
The relative HPD alignment can also be corrected using data from
the magnetic distortion measurements. 
After these steps, the alignment procedure uses the  reconstructed
Cherenkov angle of 
$\beta \approx 1$ particles to align the whole RICH detector, 
the HPD panels, and eventually
the 4 (56) spherical and 16 (40) flat
mirror segments in RICH\,1 (RICH\,2).

\begin{figure}[htbp]
\centering
\includegraphics[scale=0.4]{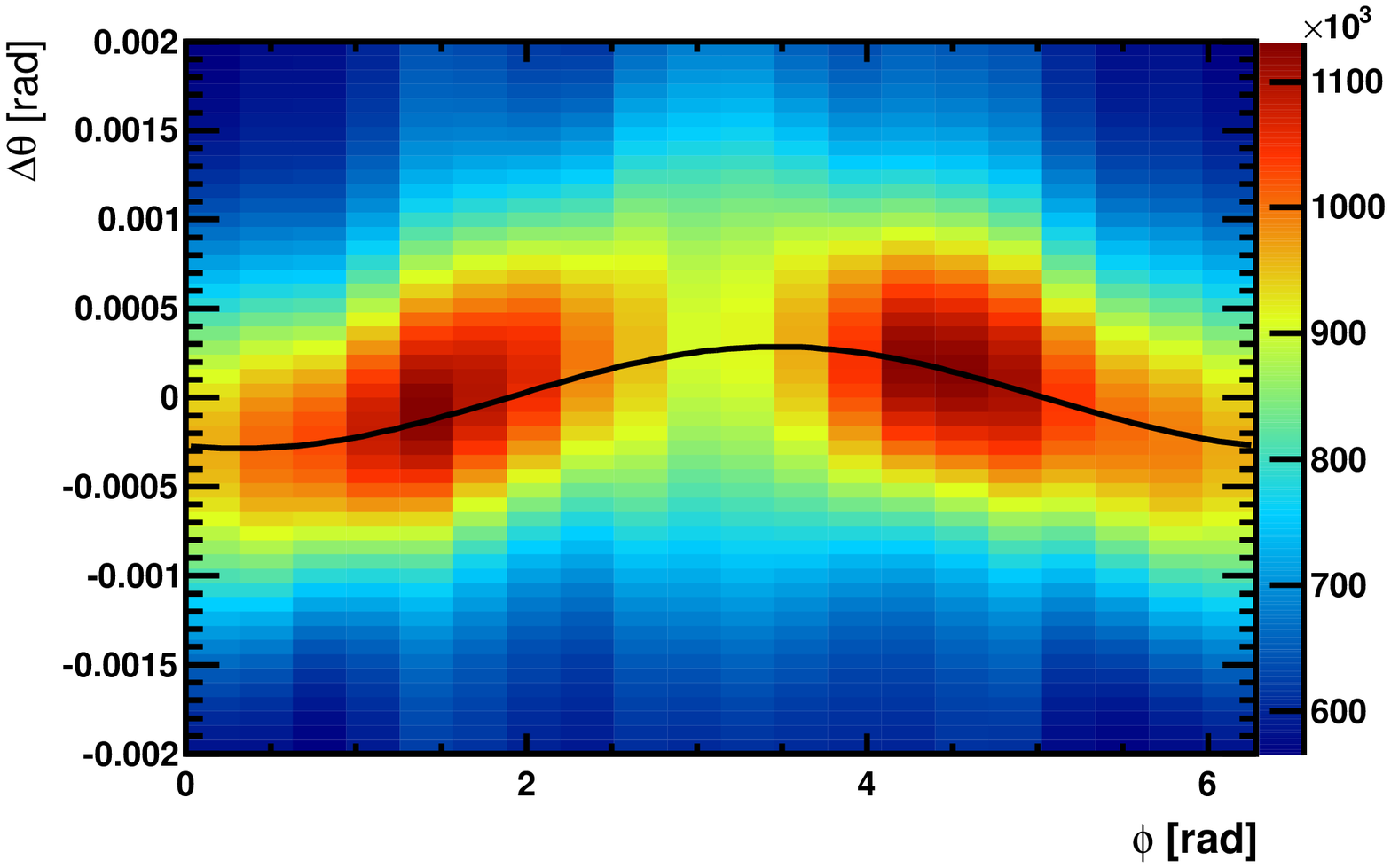}
\includegraphics[scale=0.4]{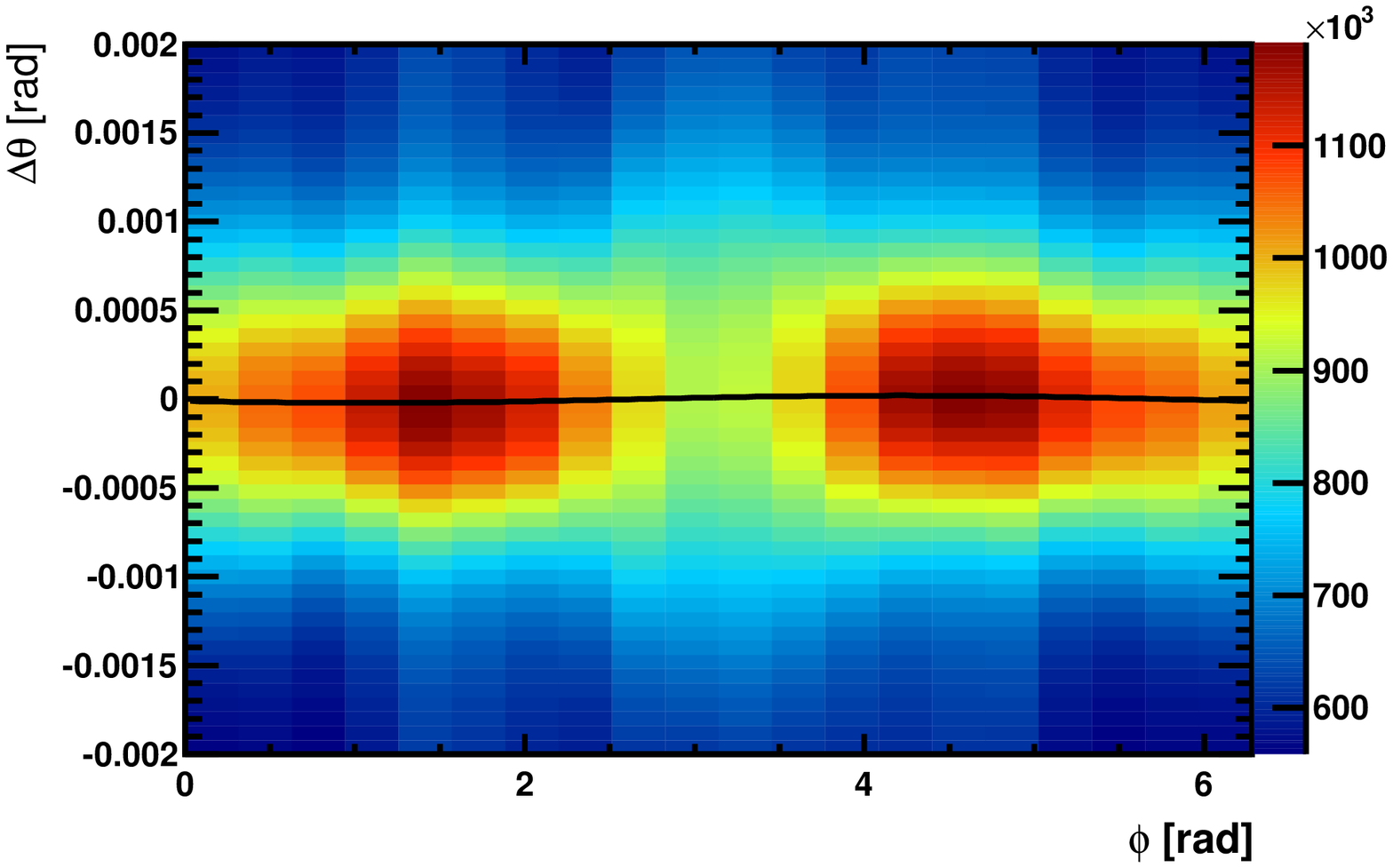}

\caption{$\Delta\theta_C$ plotted as a function of the azimuthal angle $\phi$
and fitted with $\theta_{x} \cos(\phi)
+ \theta_{y}\sin(\phi)$, for one side of the RICH\,2 detector.
The left-hand plot is prior to alignment, and 
shows a dependency of the angle $\theta_{C}$ 
on the angle $\phi$. The right-hand plot is after the alignment correction,
 and $\Delta \theta_{C}$ is uniform in $\phi$}
\label{fig:dthetaPhiSide0R2}
\end{figure}

Any misalignment of the RICH detectors with respect to the tracking 
system is observed as a shift of the track projection point
on the photodetector plane from the centre of the corresponding Cherenkov
ring. This shift 
is observed by analysing the Cherenkov angle, $\theta_{C}$, 
as a function of the azimuthal Cherenkov angle $\phi$, defined as the angle
of the pixel hit in the coordinate system of the photodetector plane,
with the projected track coordinate at the origin. 
The angle $\theta_{C}$ is independent of the angle $\phi$ for a well 
aligned detector, whilst a
misaligned system would result in a sinusoidal distribution as shown in 
Fig.~\ref{fig:dthetaPhiSide0R2}.

In practice, distributions of $\Delta\theta_C$ against $\phi$ are plotted where 
$\Delta \theta_C = \theta_{C} - \theta_{0}$
and $\theta_{0}$ is the Cherenkov angle calculated from
the momentum of the track and the refractive index of the radiator.
Any systematic shift away 
from the value $\theta_0$ is observable as a shift in
$\Delta\theta_C$.

\noindent The  $\Delta\theta_C$ distribution is then divided into slices
in $\phi$. For each slice, a one dimensional histogram of
$\Delta\theta_C$ is fitted with a Gaussian plus a second order
polynomial background and the peak of the distribution is extracted.
The mean of each slice fit is then used to fit a sinusoidal distribution given
by 
\[
\Delta \theta_C = \theta_{x} \cos(\phi) + \theta_{y}\sin(\phi).
\]

\noindent The final fit is shown in Fig. 
$\ref{fig:dthetaPhiSide0R2}$; the extracted values  
of $\theta_{x}$ and $\theta_{y}$ 
correspond to a misalignment on the HPD detector plane in the $x$ and $y$ 
direction respectively.


The alignment of the mirror segments has the extra complication that every
photon is reflected twice, and so the data must be separated into samples
which have unique 
spherical and flat mirror combinations. For this procedure, only photons
that can be uniquely associated to a given mirror pair are used.
Mirror segments are identified by considering photons to have been emitted 
at both the start and end of the gas radiators. 
If the  mirror segments reflecting the photons are the same in both cases, 
the photon trajectory is considered 
unambiguous and is used for the alignment of mirror segments.

The mirror arrangement in RICH\,1 allows for alignment using a 
sequential approach as described above, where the spherical 
mirrors are aligned first, followed by the planar mirrors. 
This is possible because photons reaching a
particular planar mirror can only be reflected from a single spherical mirror
\cite{Alignment}. 
In RICH\,2 the larger number of spherical/planar 
mirror combinations makes the use of
a sequential method impossible. 
The alignment of the RICH\,2 mirror segments is
performed by solving a set of simultaneous equations to extract all the
alignment parameters of all the mirrors. One iteration of this method 
is required to obtain the final mirror alignment.

\subsection{Refractive index calibration}
The refractive index of the gas radiators  depends 
on the ambient temperature and pressure and the exact composition of the
gas mixture. It can therefore change in time, and this 
affects the performance of the particle identification algorithms.
The ultimate calibration of the refractive index is performed using 
high momentum charged particle
tracks in such a way that the distribution of $\Delta\theta_{C}$ 
peaks at zero. 

The calibration of the refractive index of the aerogel is performed 
using tracks with momentum $p> 10$ GeV/$c$ passing through each tile.
It is found not to change as a function of time.

\subsection{Monitoring hardware}

There are additional monitoring tasks, independent from the methods described 
above.

The four spherical mirrors in RICH\,1 and 20 of the mirror segments in RICH\,2
are monitored for stability using laser beams
and cameras. For each monitored mirror there is an optic fibre with a lens to
provide a focused beam, a beam splitter, a mirror and a camera. 
The beam splitter creates two beams. 
The reference beam is incident directly onto the camera. 
The second beam is reflected to the camera via 
the monitored mirror. A comparison of the relative position of these 
light spots tracks possible movement of the mirror. 

The purity of the gas radiators is monitored by measuring the speed 
of sound in the gas. A 50 kHz ultrasonic range finder is used. 
The gas to and from the detector is monitored with a precision of 
about 1\% for a binary gas mixture. 
A gas chromatograph is periodically used for high precision measurements.
Any variation in time, after correction for temperature effects, is likely 
due to changes in the composition of the gas. 

\begin{figure}
\begin{center}
\includegraphics[angle=-90,width=0.45 \textwidth]{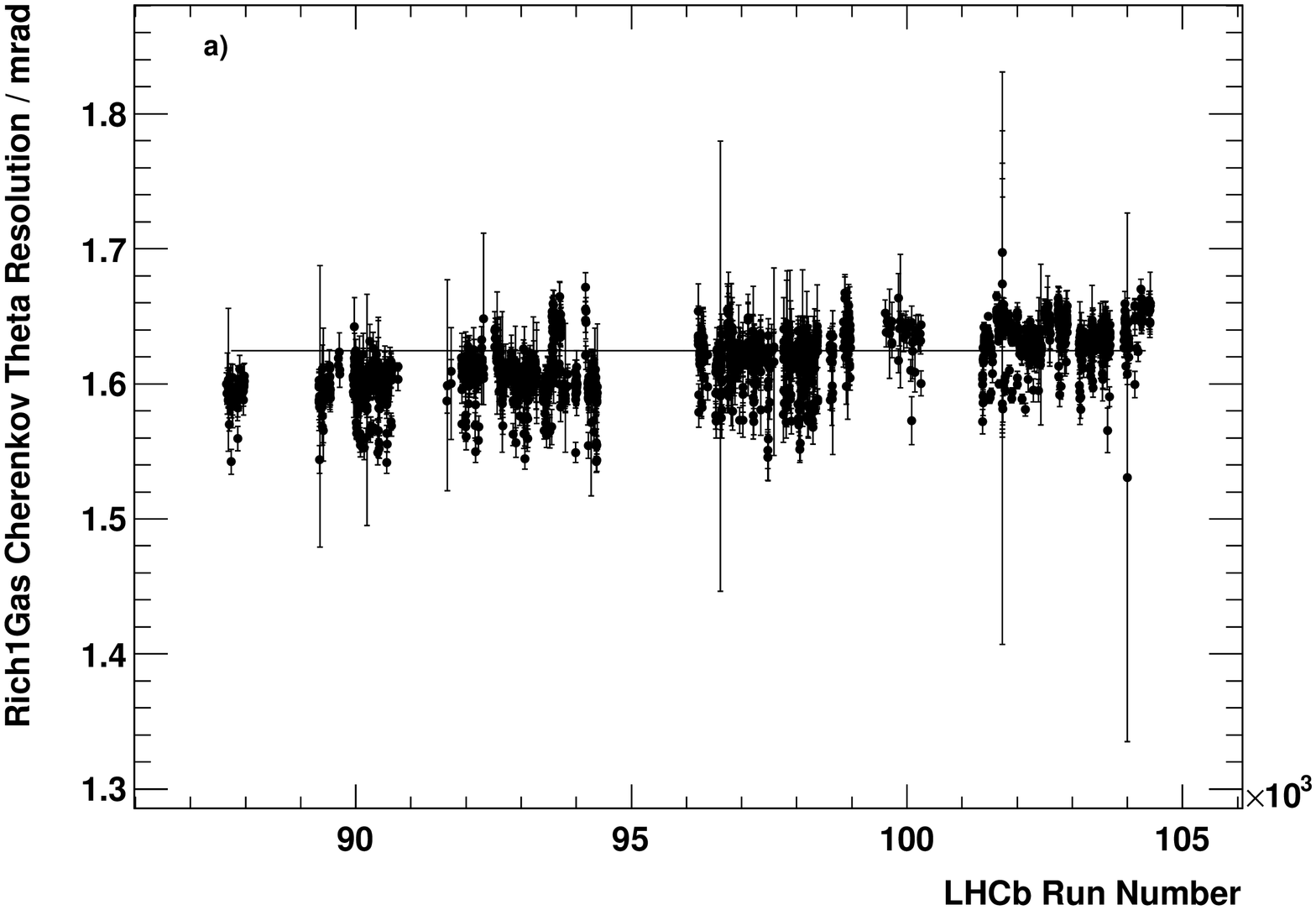}
\includegraphics[angle=-90,width=0.45 \textwidth]{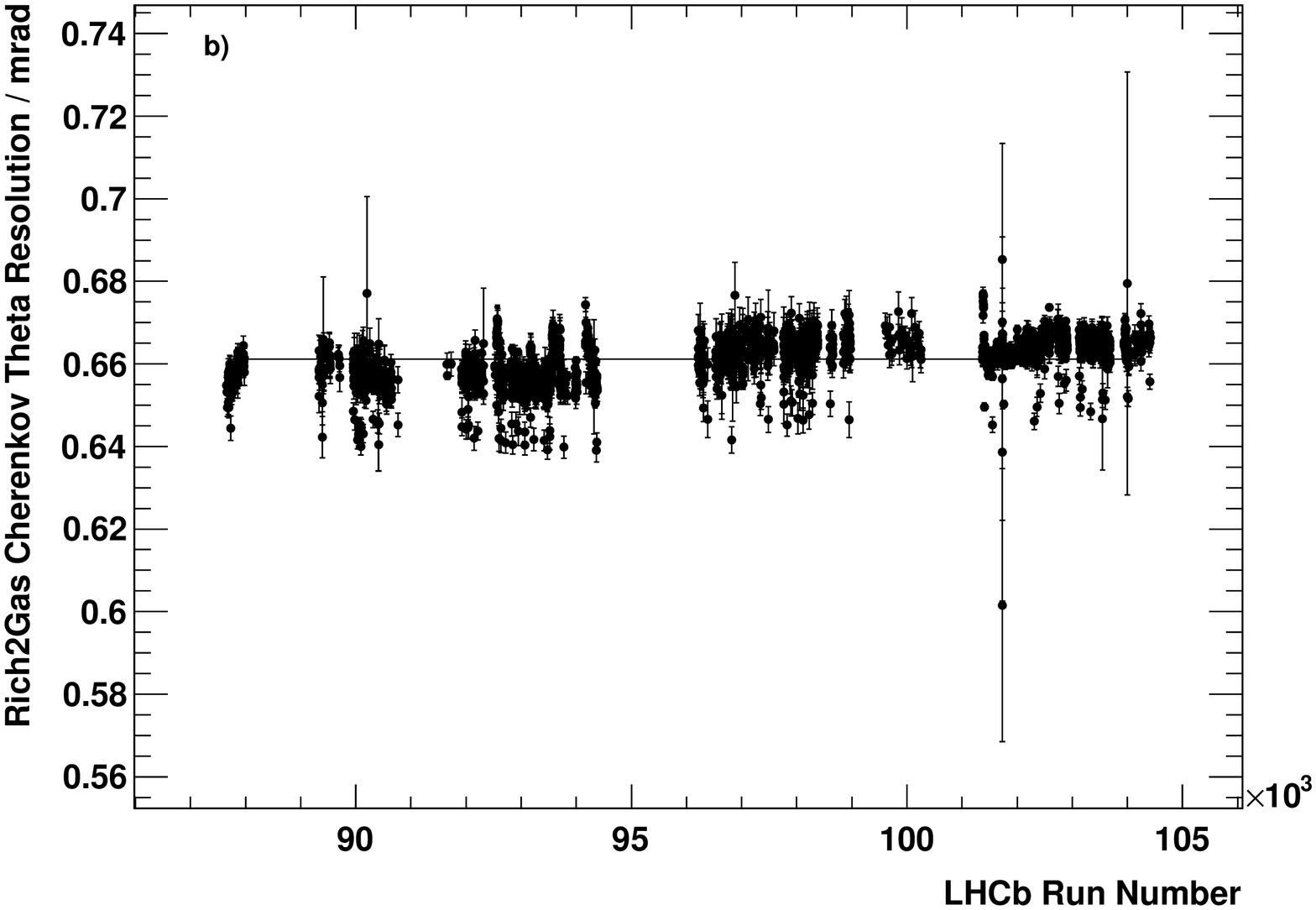}
\caption{The Cherenkov angular resolution
(c.f. Sect. \ref{sec:AngularResolution}),
 after all corrections have been applied, as a function of run number.
a) for RICH\,1 and b) for RICH\,2. 
The period of time covered on the x-axis corresponds to about 8 months of
running
}
\label{fig:res_vs_time}
\end{center}
\end{figure}

After correcting for all the parameters monitored as a function of time as
described in this section, the detector behaviour is very stable, as
shown in Fig.~\ref{fig:res_vs_time}.

\section{Performance}

\subsection{Data reconstruction}
\label{sec:Software}

The LHCb software is based on the Gaudi Framework 
\cite{GaudiA,GaudiB} which
provides a flexible and configurable C++ Object Oriented framework.
This flexibility allows the same software to be used in a
variety of different RICH applications, ranging from the online monitoring, 
the utilization
of the RICH in the final stages of the higher level trigger, and providing 
the full offline
event reconstruction. This section describes the
processing steps of the RICH data. 

\subsubsection{HPD data reconstruction}
\label{sec:Software:HPDData}

The first stage of the data processing chain is to decode the raw data,
as read out from the detector, to offline storage. 
This produces a list of the HPD pixels that have been hit in each event.
The next step is to apply various data cleaning algorithms to the list of
active pixels for each HPD. 
HPD data are rejected if the HPD occupancy, which on average is $\sim$ 1\%,
exceeds a tuneable maximum value of 20\%, to exclude excessively large 
events.

Finally, the position of the photon hit is reconstructed on the HPD plane.
This procedure corrects for the alignment of the HPDs 
within the LHCb detector, the electrostatic focusing parameters of the HPD 
tubes, and the corrections for the magnetic field (Sect. \ref{sec:MDMS}). 

\subsubsection{Cherenkov photon candidate reconstruction}
\label{sec:Software:PhotonReco}

The tracking
system of LHCb provides detailed coordinate information on the passage of 
charged particles through the LHCb spectrometer, and with this information the 
trajectory of each particle
through the three RICH radiator volumes can be determined. 
This allows the  computation of an assumed emission point of the photon 
candidates for each track. 
As the exact emission point of each photon is unknown (and can be anywhere 
along the particle trajectory through the radiator), the mid-point of the
trajectory in the radiator is taken.

The candidate photons for each track are determined by combining
the photon emission point with the measured hit positions of the photons.
Once the photon candidates have been
assigned, quantities such as the Cherenkov angle $\theta_C$, 
can be computed.
A full analytical solution of the RICH optics is used, which 
reconstructs the trajectory of the photon 
 through the RICH optical system, taking into account the
knowledge of the mirror and HPD alignment 
\cite{forty}.

\subsection{Cherenkov angle resolution}
\label{sec:AngularResolution}

\begin{figure}
\begin{center}
\includegraphics[width=0.45 \textwidth]{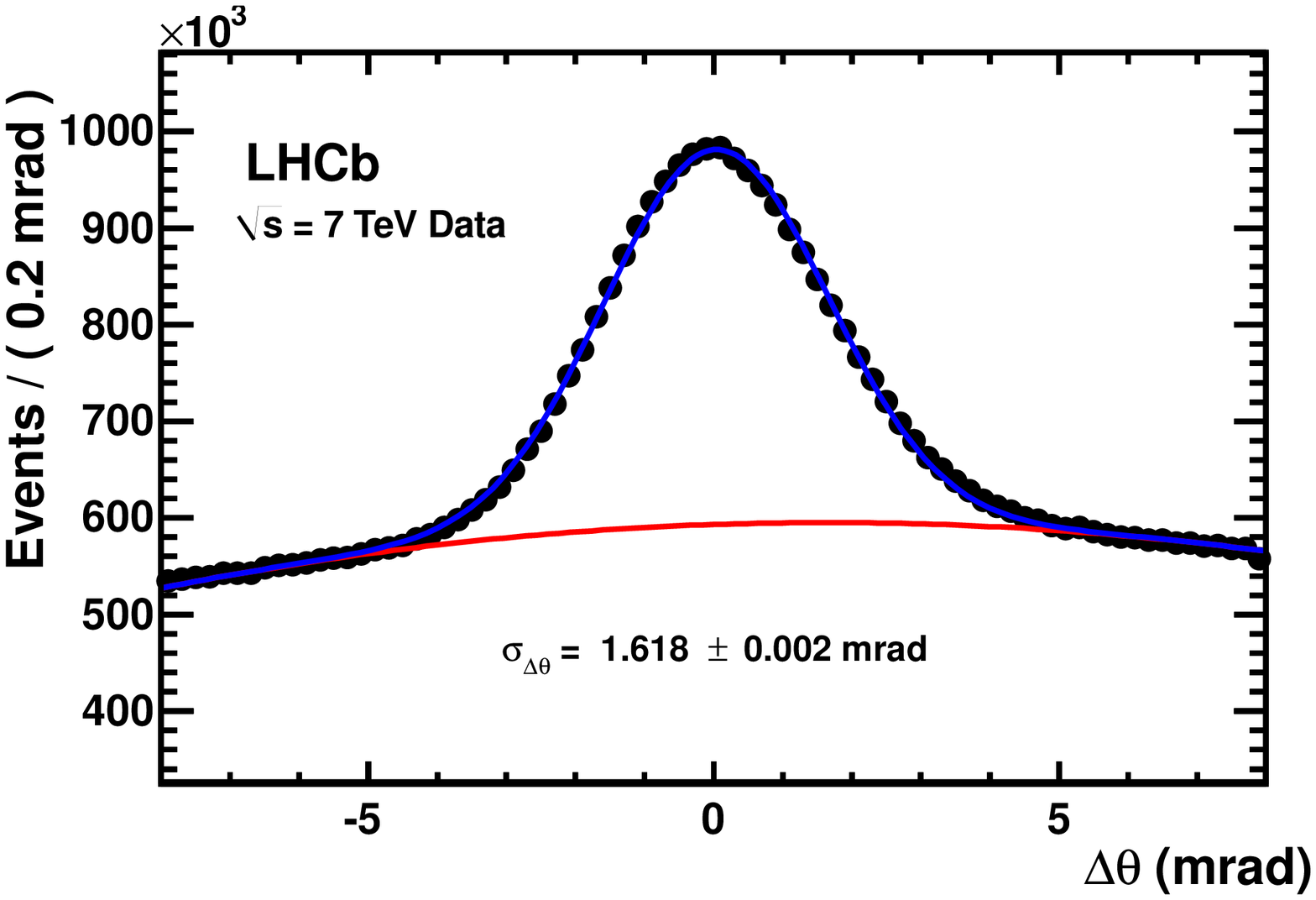}
\includegraphics[width=0.45 \textwidth]{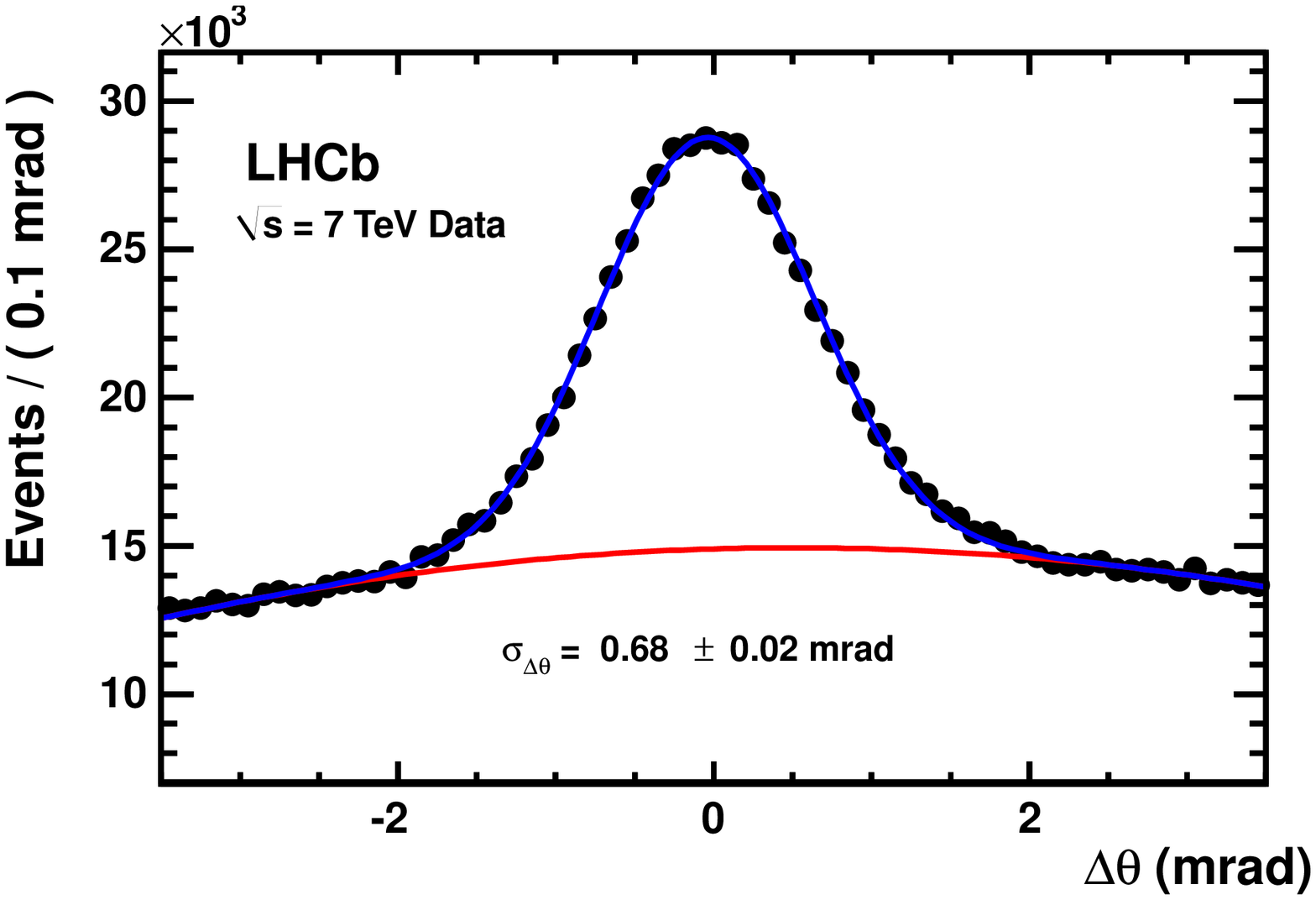}
\caption{Single photoelectron resolution for the RICH\,1 (left) and RICH\,2
 (right) gases, as measured in data for high momentum charged particles.
The red line describes the background as determined from the fit using a 
polynomial function together with the Gaussian for the signal 
}
\label{fig:R1angle_res}
\end{center}
\end{figure}

The distribution of $\Delta\theta_C$, calculated for each photon with respect
 to the measured track, is shown for the RICH\,1 and RICH\,2 gas radiators
in Fig.~\ref{fig:R1angle_res}
after the alignment and calibration procedures have been performed.

By fitting the distribution with a Gaussian plus a polynomial background,
the Cherenkov angle resolution is determined to be 
$1.618 \pm 0.002$ mrad 
for C$_{4}$F$_{10}$  and
$0.68 \pm 0.02$ mrad for CF$_{4}$.
 These values are in reasonable agreement with the expectations from simulation 
\cite{geant4} of $1.52 \pm 0.02$ mrad and $0.68 \pm 0.01$ 
mrad in RICH\,1 and RICH\,2, respectively.

\begin{figure}[htb!]
\begin{center}
\includegraphics[width=0.6\textwidth]{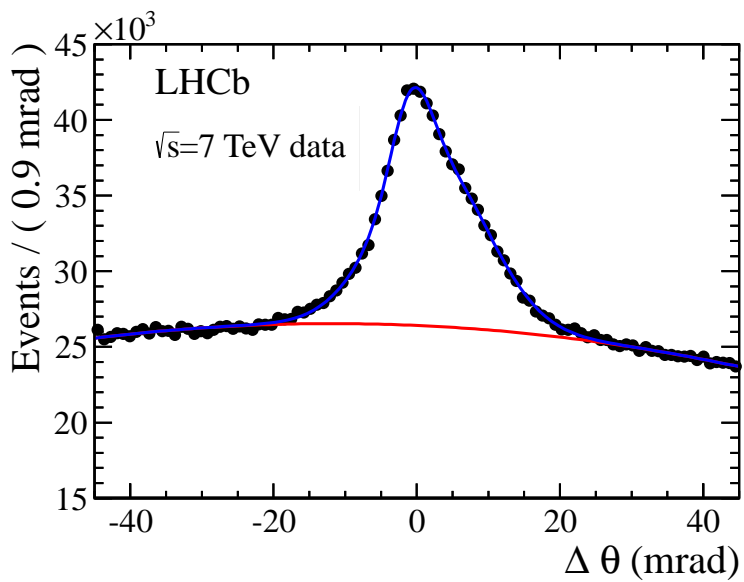}
\caption{Single photoelectron resolution for the aerogel as measured in 2011 
data with the pp$\rightarrow$pp$\mu^+\mu^-$ events.  
The red line describes the background as determined from the fit using a 
polynomial function together with two Gaussians for the signal 
}
\label{fig:aerogel_res}
\end{center}
\end{figure}

The performance of the aerogel radiator has been studied
with data collected in 2010 and 2011.
The data have been first used to calibrate the refractive indices of 
individual tiles.
Figure~\ref{fig:aerogel_res} shows the deviation $\Delta\theta_C$
 in the four aerogel tiles located around the beampipe, 
which cover more than 90\% of the acceptance. 
The $\Delta\theta_{C}$ distribution of the
photons is measured using good quality tracks  with momentum above 
10 GeV$\slash c$.
 The peak is not symmetric, and the $\sigma$ from the
FWHM gives an average value of about 5.6~mrad (the events used for this estimate
are all pp collisions, not the ones used in Fig.~\ref{fig:aerogel_res}).
This value is about a factor of 1.8 worse than the simulation. 
This discrepancy is, at least partially, explained by
the absorption by the very porous aerogel structure of the
 C$_{4}$F$_{10}$ with which it is in contact.

A new aerogel enclosure which isolates the aerogel from
the C$_4$F$_{10}$ gas in RICH\,1 is installed for the 2012 running.


\subsection{Photoelectron yield}
\label{sec:Npe}


\begin{figure}[b!]
 \centering
 \includegraphics[width=0.6\textwidth]{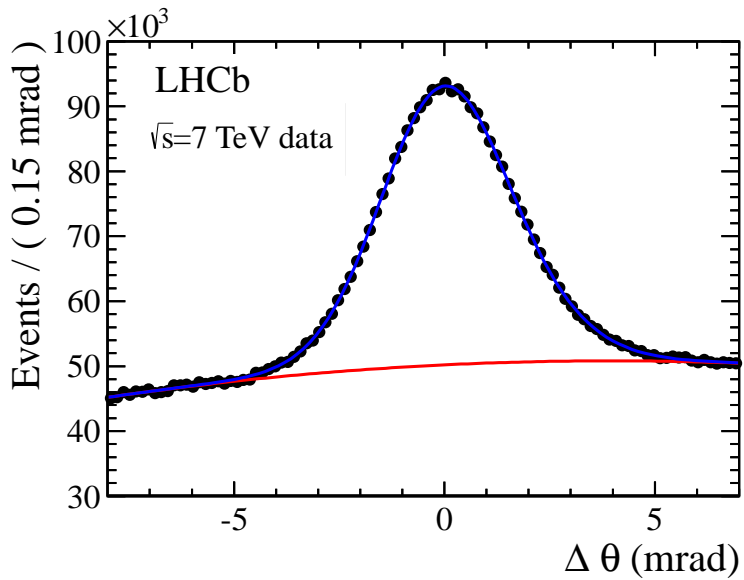}
 \caption{Distribution of $\Delta \theta_C$ for C$_4$F$_{10}$. 
This plot is produced from kaons and pions from tagged 
D$^0 \rightarrow$ K$^- \pi^+$ decays in data selected with the criteria 
described in the text}
 \label{fig:dtheta}
\end{figure}

The photoelectron yield N$_{\rm pe}$ is measured for two categories of RICH event: 
one, referred to as a \emph{normal} event, is representative of nominal RICH 
running conditions during LHCb physics data collection; the other, referred to
 as an \emph{ideal} event, is a special event type with very low 
photoelectron backgrounds and clean tracks with full, unobstructed Cherenkov 
rings.

The \emph{normal} event category uses an unbiased (in that the RICH detectors are 
not used in the selection) track sample composed of 
kaons and  pions originating
from the decay 
D$^0 \rightarrow$ K$^- \pi^+$, where the D$^0$ is selected from  
D$^{*+}\rightarrow $D$^0 \pi^+$ decays.
The kaons and pions are required to have track momenta
 $p_{\rm K}>9.8$ GeV/c and $p_\pi>5$ GeV/c in the aerogel;
$p_{\rm K}>37$ GeV/c and $p_\pi>30$ GeV/c in C$_4$F$_{10}$, 
and $p_{\rm K}>74.8$ GeV/c,  $p_\pi>40.4$ GeV/c in CF$_4$. 
These cuts ensure that all tracks have an expected Cherenkov angle close to
saturation ($\beta\approx 1$).

The track sample of the \emph{ideal} event category is composed of muons 
selected from pp$\rightarrow$pp$\mu^+ \mu^-$ events. 
The events are required not to have a visible primary vertex. 
The track momentum selection criteria of the muons is the same as for pions in 
the \emph{normal} event category. 
A cut was applied on the track geometry, such that at least half of the  
Cherenkov cone associated to the track projects onto the HPD pixels.
This selection avoids losses owing to the cone intersecting with 
the beampipe, or projecting onto the region outside the HPD acceptance and 
the gaps between the HPDs.

 N$_{\rm pe}$ is measured by fitting the  $\Delta \theta_C$
distributions of the photoelectrons.
For each selected charged particle track, photoelectron hits that lie within 
a $\Delta \theta_C$ range of $\pm 5\sigma$, where $\sigma$ is the Cherenkov 
angle resolution, are retained.
Photoelectrons that are correctly associated with a track peak around 
$\Delta \theta_C = 0$ and are distributed as a Gaussian,
while those from other tracks and background sources
form a non-peaking background, as shown in Fig.~\ref{fig:dtheta} obtained from C$_4$F$_{10}$. 

\begin{figure}[ht!]
  \centering
  \includegraphics[width=0.45\textwidth]{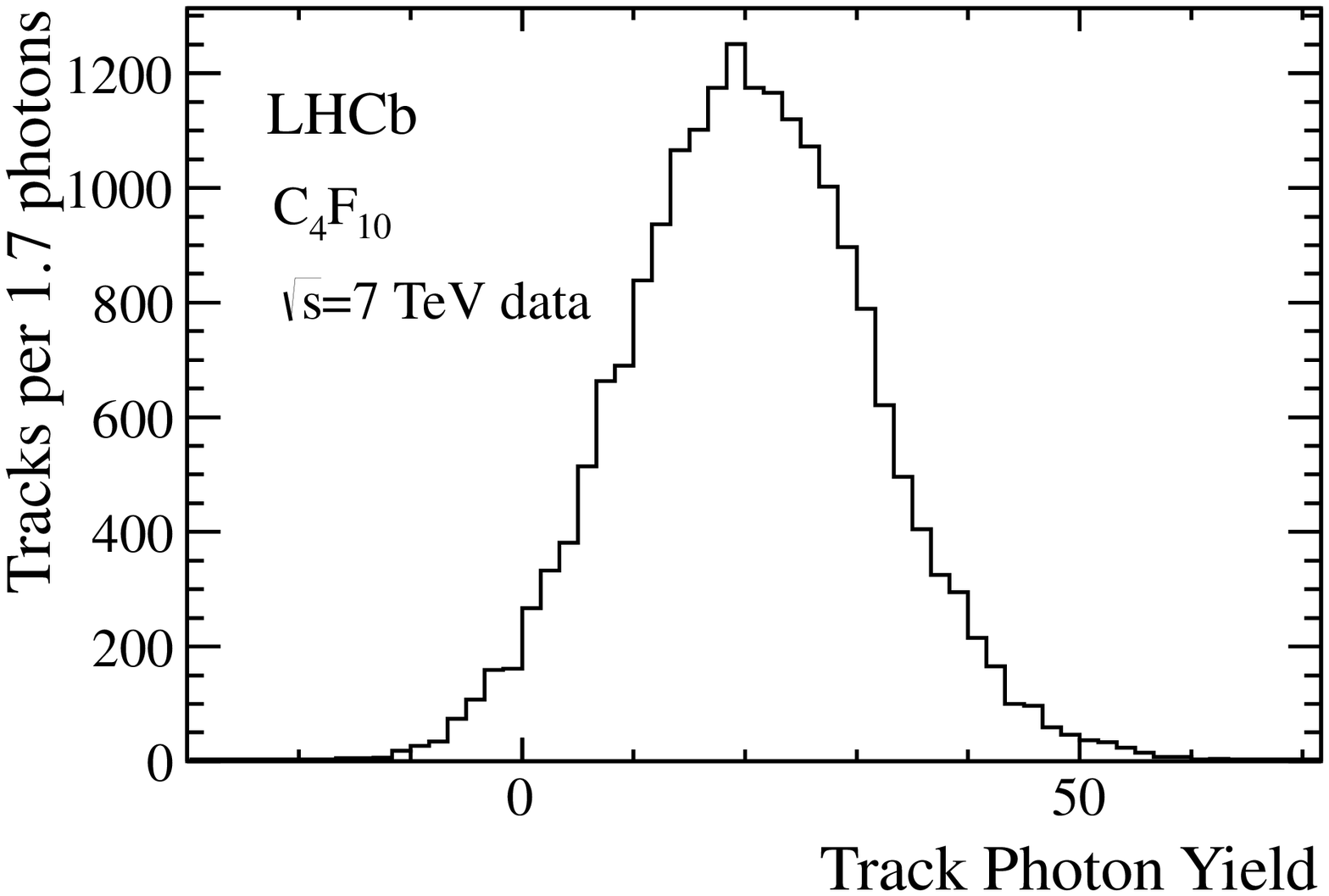}
  \includegraphics[width=0.45\textwidth]{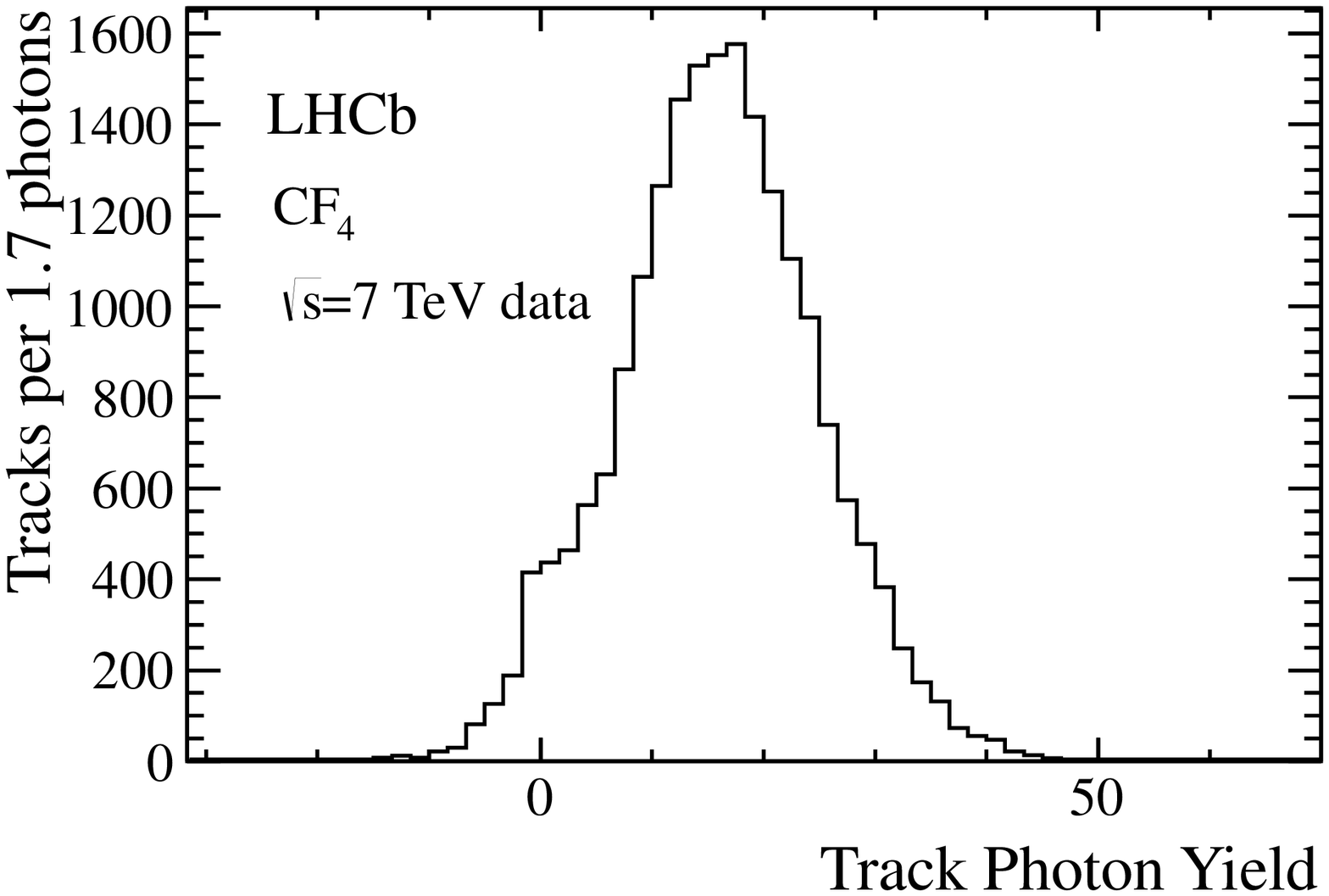}
  \caption{Individual track photon yield distributions for the 
C$_4$F$_{10}$ (left) and CF$_4$ (right) radiators. The plot is produced 
from kaons and pions from tagged D$^0 \rightarrow$ K$^- \pi^+$ decays in 
data selected with the criteria described in the text}
  \label{fig:yield_distros}
\end{figure}

An initial fit is performed on the $\Delta \theta_C$ distribution aggregated 
from all the selected tracks, using a probability density function (PDF) 
composed of a Gaussian signal over a quadratic background. 
The $\Delta \theta_C$ distribution of each individual track is then fitted with
 a Gaussian signal over a linear background PDF, with the mean of the Gaussian 
fixed at 0 and the width fixed to that obtained from the fit to the aggregated
 $\Delta \theta_C$ distribution. 
The individual track N$_{\rm pe}$ is then taken as the number of photoelectron 
candidates under the Gaussian shape. 
The overall value for N$_{\rm pe}$ is taken as the mean of the distribution of 
the track N$_{\rm pe}$, with the error corresponding to the standard error on the mean.
Figure ~\ref{fig:yield_distros} shows the data distributions at the basis of the
measurement.

The validity of the N$_{\rm pe}$ calculation method was assessed using 
simulated samples of 
D$^{*+} \to$ D$^{0}$(K$^{-}\pi^{+})\pi^{+}$ decays. 
The same selection criteria were applied as in data and 
in addition the track geometry selection was applied with the same 
criteria as for \emph{ideal} RICH events, to allow a like-for-like comparison 
between simulation and \emph{ideal} data events.
The calculated value for N$_{\rm pe}$ was compared to the true photoelectron yield,
 which was taken by counting the number of photons associated to each track by 
the simulation and then taking the average over all tracks. 

To allow a like-for-like comparison of the true and calculated N$_{\rm pe}$ values 
in the simulation study, events were required to have less than 50 hits in the 
Scintillator Pad Detector (SPD) \cite{Detector}, which gives an approximate 
measurement of the charged track multiplicity in the event. 
It has been observed that the measured N$_{\rm pe}$ is lower for high track 
multiplicity events, which have high HPD occupancies (more than $20\%$ in the 
central HPD's in RICH\,1 for events with $>500$ charged tracks).
This results in a loss 
of detected photoelectrons, because instances where multiple photons hit the 
same pixel result in only one photoelectron hit due to the binary HPD readout. 
This suppression of N$_{\rm pe}$ was not observed when an analog HPD readout was 
emulated in the simulation.

Table \ref{table:results} shows the results of the analysis performed on real 
data and on the simulation. 
In the simulated data, the calculated and true values of N$_{\rm pe}$ are in 
good agreement for all the radiators.
This validates the accuracy of the yield calculation. 
The N$_{\rm pe}$ values for the \emph{ideal} events are 
less than those from the simulation sample. 
The \emph{normal} events have values of N$_{\rm pe}$ that are less than those 
for \emph{ideal} events. This is mainly due to the higher charged track 
multiplicities of the \emph{normal} events, reducing the 
N$_{\rm pe}$, and the track geometry cut that is applied to the \emph{ideal}  
events increasing their N$_{\rm pe}$ yield. 
The aerogel N$_{\rm pe}$ data values have a large uncertainty due to the large 
background in the $\Delta \theta_C$ distributions and the additional 
uncertainty in the shape of the signal peak.

The photoelectron yields are lower than those predicted by the simulation:
however, there is evidence that the yield in data can be increased by a
few percent in RICH\,1 by retuning the setting of the HPD readout chip.
This retuning was found necessary for all HPDs by the fact that the trigger 
rate went 
up significantly during 2011 running, resulting in a readout inefficiency. 
Furthermore, the detailed description of the detector in the 
simulation needs continous retuning, especially for a RICH detector 
where the Cherenkov photons must interact with many detector elements. 
It must be stressed however, that the smaller yield measured in
data does not have a consequence on the final particle identification 
performance, as described in Sect.~\ref{sec:PID}.


\begin{table}[ht]
\centering
\begin{tabular}{|c| c| c | c |c|}
\hline\hline
 & \multicolumn{2}{|c|}{N$_{\text{pe}}$ from data} & \multicolumn{2}{|c|}{N$_{\text{pe}}$ from simulation} \\ \hline
Radiator &  tagged D$^0 \rightarrow$ K$^- \pi^+$ &pp$ \rightarrow$ pp $\mu^+ \mu^-$ & Calculated $N_{\text{pe}}$ & true $N_{\text{pe}}$ \\ [0.5ex]
\hline
Aerogel & $5.0 \pm 3.0$  & $4.3 \pm 0.9$& $8.0 \pm 0.6$ & $6.8 \pm 0.3$ \\ \hline
C$_4$F$_{10}$  & $20.4 \pm 0.1$ & $24.5 \pm 0.3$& $28.3 \pm 0.6$ & $29.5 \pm 0.5$ \\ \hline
CF$_4$ & $ 15.8 \pm 0.1$  & $ 17.6 \pm 0.2$& $22.7 \pm 0.6$ & $23.3 \pm 0.5$ \\ 
\hline
\end{tabular}
\caption{Comparison of photoelectron yields (N$_{\rm pe}$) determined from
D$^* \rightarrow$D$^0\pi^+$ decays in simulation and data, and 
pp $\rightarrow$ pp $\mu^+ \mu^-$ events in data, using the selections and 
methods described in the text }
\label{table:results}
\end{table}

\section{Particle identification performance}
\label{sec:Performance}

Determining the performance of the RICH Particle IDentification (PID),
both during and after data taking,
is particularly important for analyses that exploit RICH PID, for which 
knowledge of efficiency and  misidentification rates
  are required.
 Moreover, it enables comparison with expectations and provides a benchmark
  against which to compare the effectiveness of alignment and calibration
  procedures.


This section provides a description of the PID algorithms and the
 performance obtained following
analysis of data from the first LHC runs.

\subsection{Particle identification algorithms}
\label{sec:Software:PID}

\begin{figure}
\centering
\includegraphics[width=0.4 \textwidth]{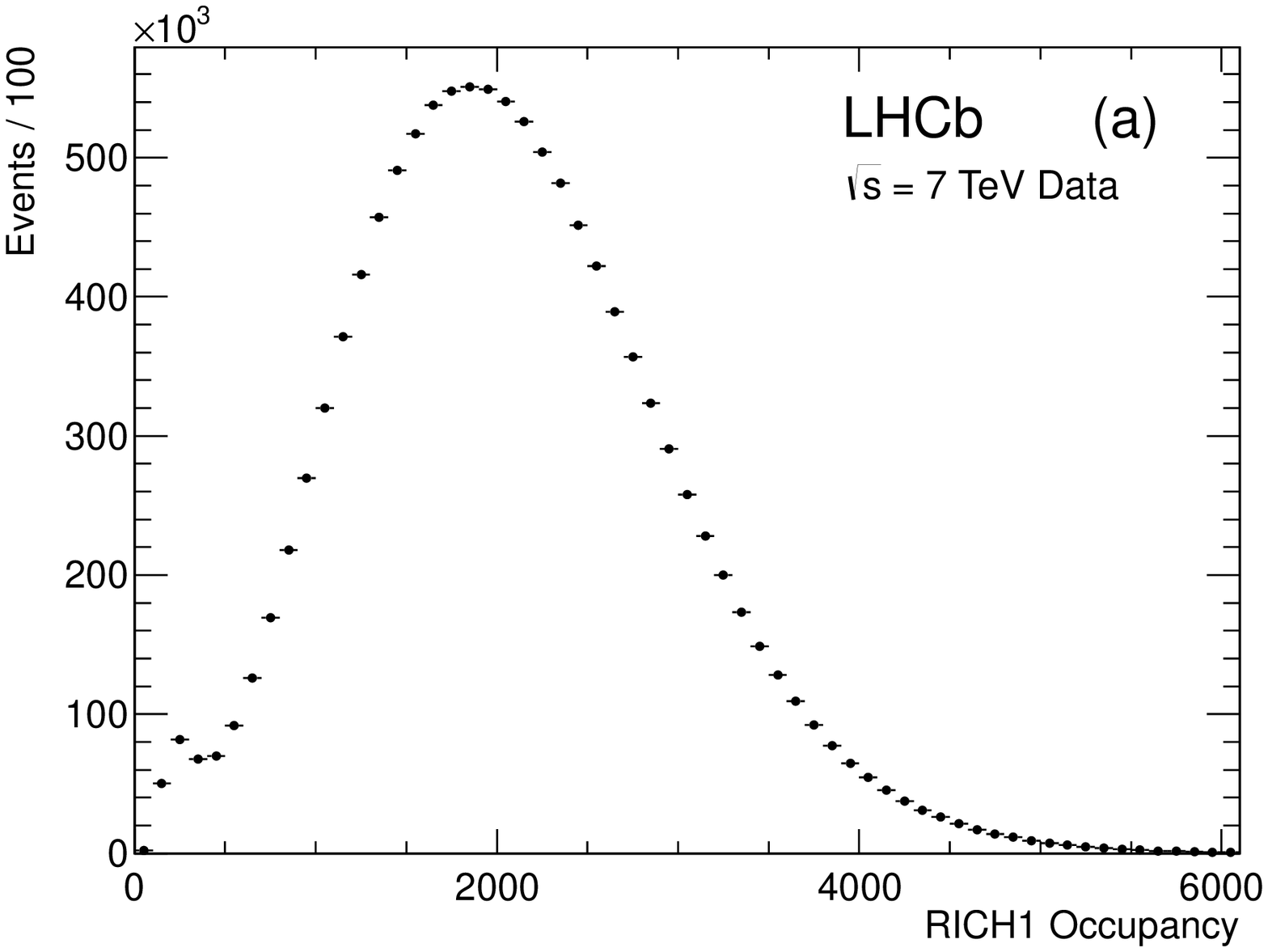}
\includegraphics[width=0.4 \textwidth]{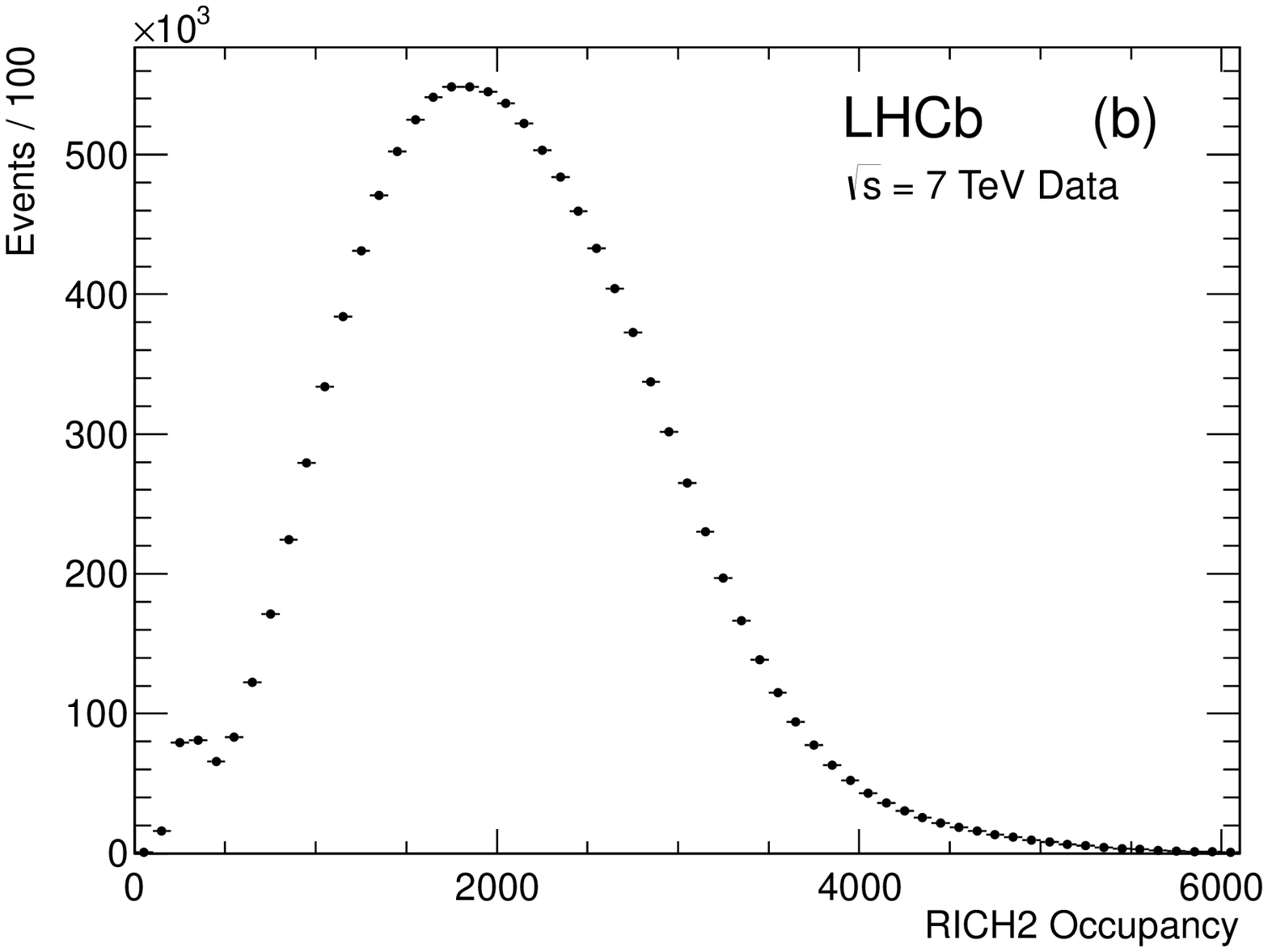}
\includegraphics[width=0.4 \textwidth]{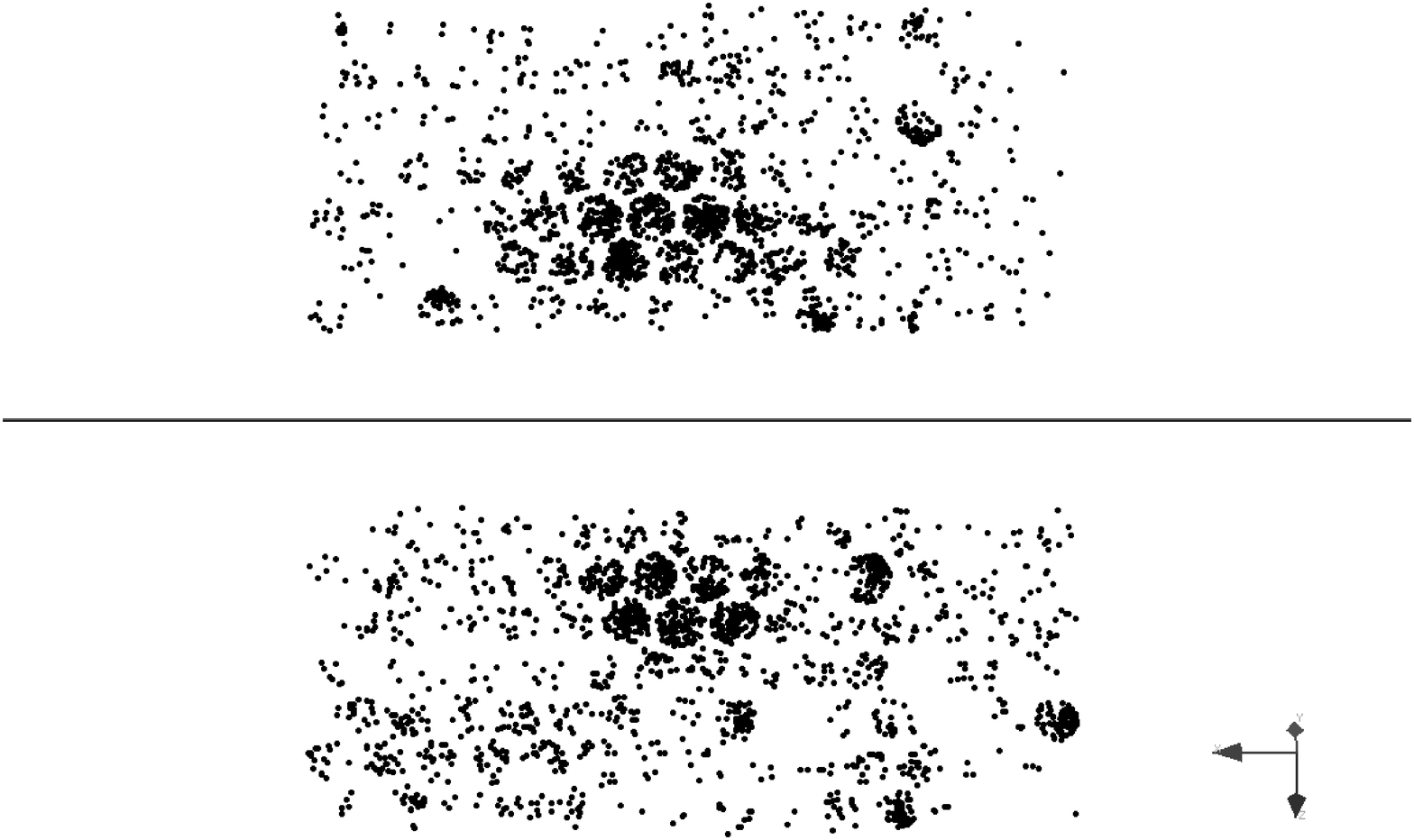}
\includegraphics[width=0.4 \textwidth]{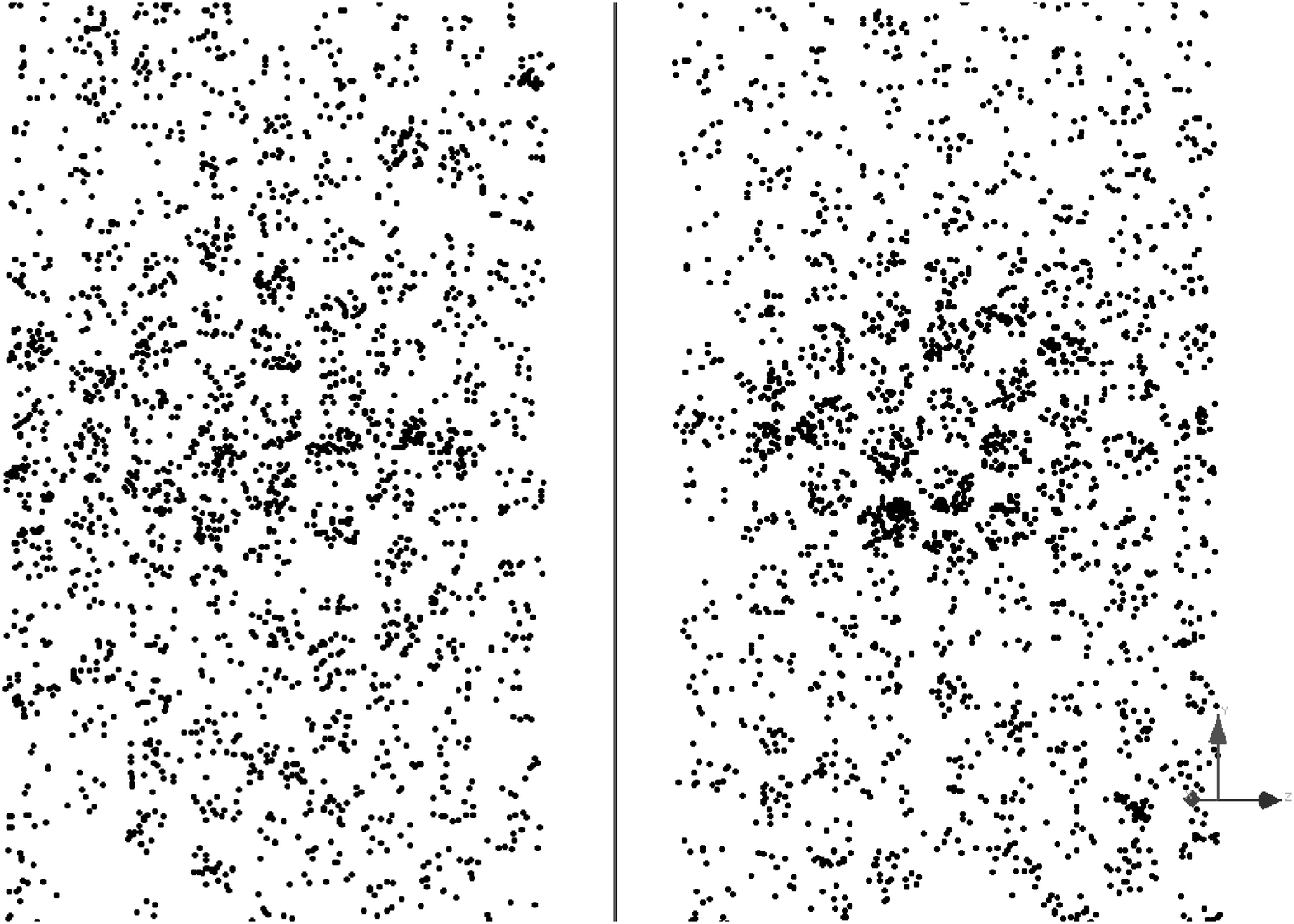}
\caption[Event Displays]{Distribution of the number of pixel hits per event in 
(a) RICH\,1 and (b) RICH\,2. An example of a typical LHCb event as seen by 
the RICH detectors, is shown below the distributions. 
The upper/lower HPD panels in RICH\,1 and the left/right panels in
  RICH\,2 are shown separately}
\label{fig:eventdisplay}
\end{figure}

In order to determine the particle species for each track,
 the Cherenkov angle information must be
combined with the track momentum measured by the tracking system,
 as described in Sect.~\ref{sec:Software:PhotonReco}.
The RICH detectors operate in a high occupancy environment, as shown in
 Fig.\ref{fig:eventdisplay}.
To reconstruct such events efficiently, an overall event log-likelihood 
algorithm is employed, where all tracks in the event and in both RICH 
detectors are considered simultaneously 
\cite{forty}.
This allows for an optimal treatment of tracks where Cherenkov
cones overlap.

Since the most abundant particles in pp collisions are pions, the
likelihood minimisation procedure starts by assuming all particles are pions.
The overall event likelihood, computed from the distribution of photon hits,
the associated tracks and their errors,
is then calculated for this set of hypotheses.
Then, for each track in turn, the likelihood is recomputed changing
the mass hypothesis to e, $\mu$, $\pi$, K and proton, whilst leaving all other
hypotheses unchanged.
The change in mass hypothesis amongst all tracks that
gives the largest increase in the event likelihood is identified, and the mass
hypothesis for that track is set to its preferred value.
This procedure is then
repeated until all tracks have been set to their optimal hypotheses,
and no further improvement in the event likelihood is found.

The procedure described above is CPU intensive for a large number of
tracks and HPD pixels, since the number of likelihood
calculations increases exponentially with the number of tracks.
In order to counter this, some modifications are made to the minimisation
procedure to limit the number of combinations, whilst still converging
on the same global solution.
During the search for the track with the
largest improvement to the event likelihood, the tracks are sorted
according to the size of their likelihood change from the previous
step, and the search starts with the track most likely to change its hypothesis.
If the improvement in the likelihood for the first track is
above a tuneable threshold, the search
is stopped and the hypothesis for that track is changed.
Secondly, if a track shows a clear preference for one mass hypothesis, then
once that track has been set to that hypothesis, it is removed in
the next iterations.
These modifications to the likelihood minimisation dramatically
reduce the CPU resources required.

The background contribution to the event likelihood is determined
prior to the likelihood algorithm described above.
This is done by comparing the expected signal in each HPD,
due to the reconstructed tracks and their assigned mass hypothesis, to
the observed signal. Any excess is used to determine the background
contribution for each HPD and is included in the likelihood calculation.

The background estimation and likelihood minimisation algorithms can
be run multiple times for each event.
In practice it is found that only two
iterations of the algorithms are needed to get convergence.
The final results of the particle identification are differences in the
log-likelihood values
 ${ \Delta \log \mathcal{L}}$,
 which give for each track the change in the overall
event log-likelihood when that track is changed from the pion hypothesis 
to each of the electron, muon, kaon and proton hypotheses.
These values are then used to identify particle types.

\subsection{Performance with isolated tracks}

A reconstructed Cherenkov ring will
generally overlap with several others.
Solitary rings from \emph{isolated} tracks
provide a useful test of the RICH performance, since the reconstructed 
Cherenkov angle can be uniquely predicted.
A track is defined as \emph{isolated} when its Cherenkov ring does not
overlap with any other ring from the same radiator.

\begin{figure}
\begin{center}
  \includegraphics[width=0.60\textwidth]{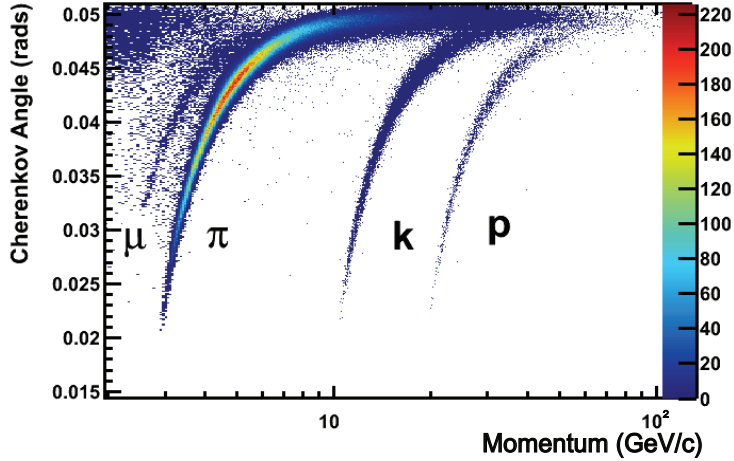}
  \caption{Reconstructed Cherenkov angle as a function of track momentum in the
 $\rm C_{4}F_{10}$ radiator
}

  \label{fig:CAngleVMom}
\end{center}
\end{figure}

Figure~\ref{fig:CAngleVMom} shows the Cherenkov angle as a 
function of particle momentum using information from the C$_4$F$_{10}$ 
radiator for isolated tracks selected in data ($\sim 2\%$ of all tracks). 
As expected, the events are distributed into distinct 
bands according to their mass. 
Whilst the RICH detectors are primarily used for hadron identification, it is 
worth noting that a distinct muon band can also be observed.

\subsection{PID calibration samples}

In order to determine the PID performance on data, high statistics samples of
genuine K$^{\pm}, \pi^{\pm}$, p and $\bar{\rm p}$ tracks are needed.
The selection of such control samples must be independent of PID information,
 which would otherwise bias the result.
The strategy employed is to reconstruct,
through purely kinematic selections independent of RICH information,
exclusive decays of particles copiously produced and reconstructed at LHCb.

\begin{figure}[h]
  \begin{center}
    \subfigure[]{\label{fig:Ks0Mass}
      \includegraphics[width=0.4 \textwidth]{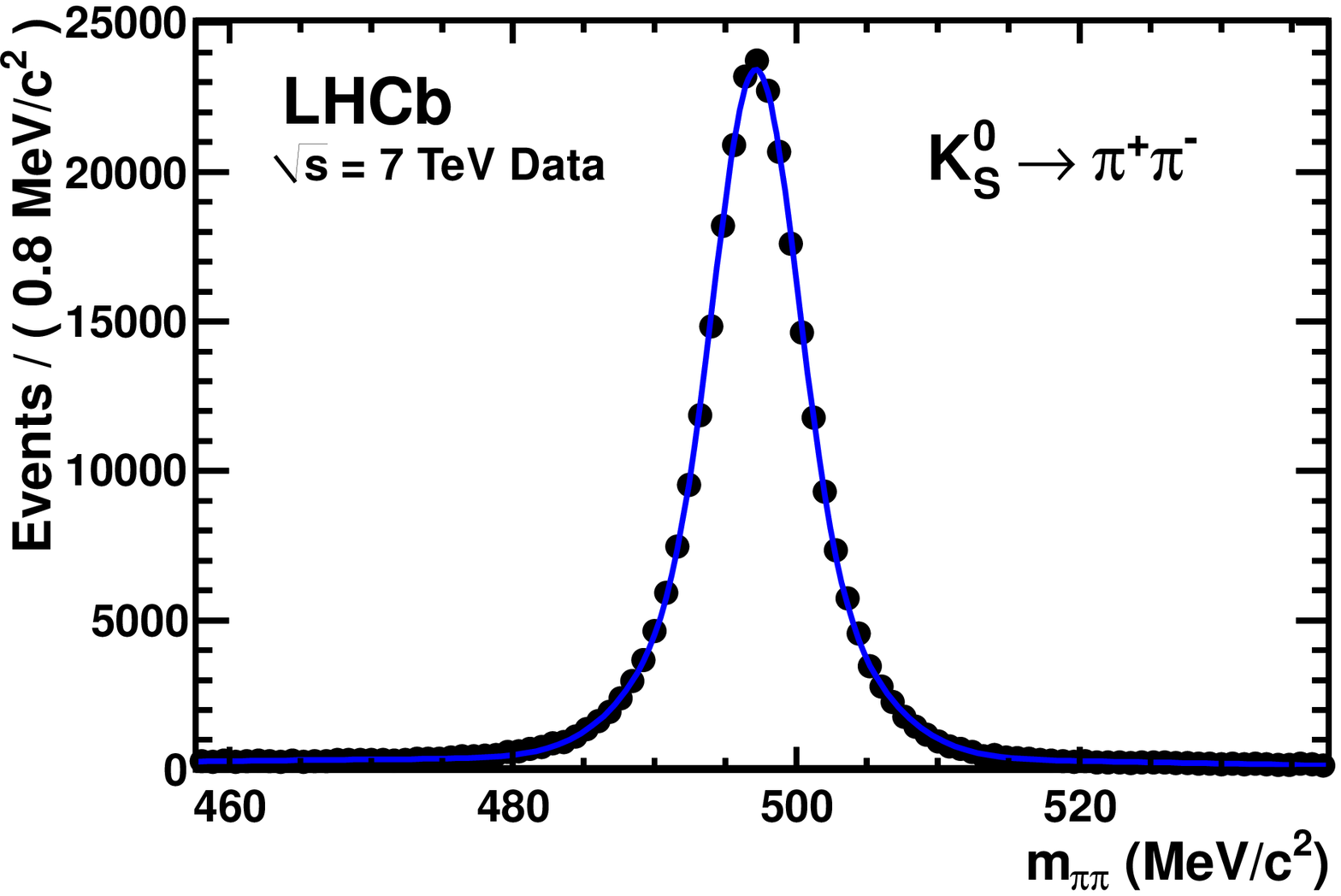}}
    \subfigure[]{\label{fig:Lam0Mass}
      \includegraphics[width=0.4 \textwidth]{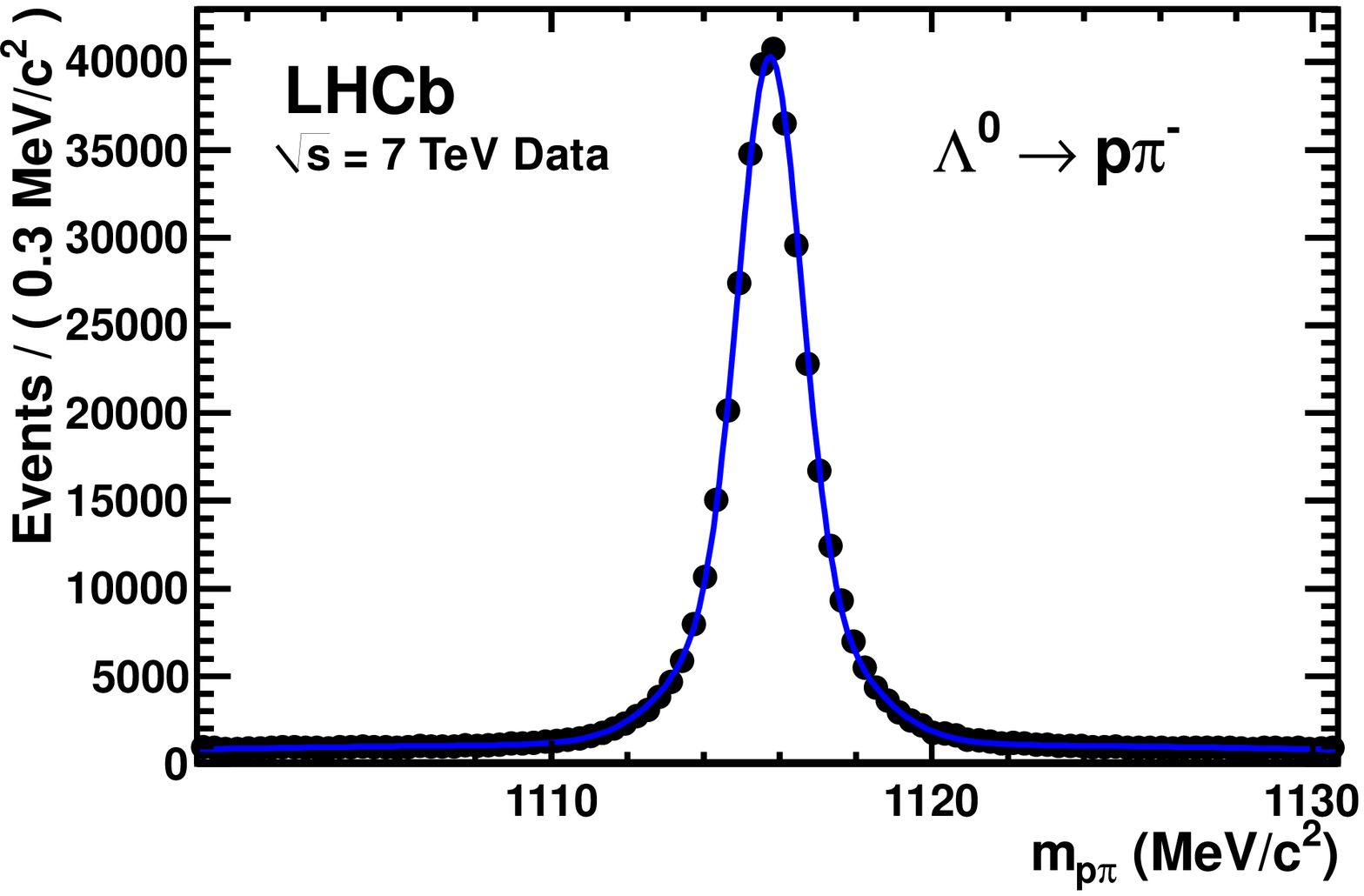}}
    \subfigure[]{\label{fig:D0Mass}
      \includegraphics[width=0.4 \textwidth]{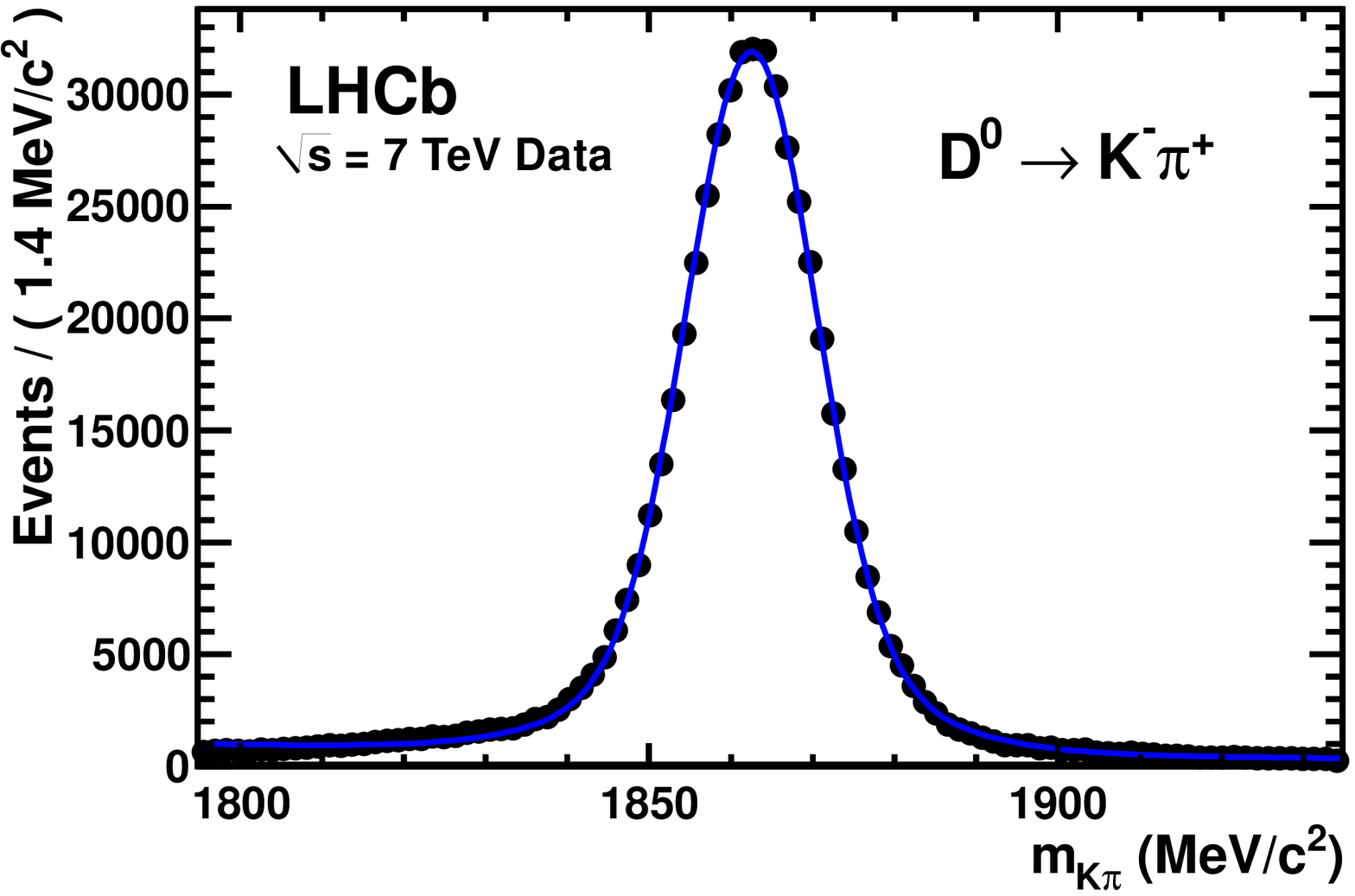}}
    \caption{Invariant mass distributions of the (a) K$^{0}_{S}$, 
(b) $\Lambda$ and (c) D$^{0}$ calibration samples. 
The best fit probability-density-function (pdf), describing both background 
and signal, is superimposed in blue}
    \label{fig:MassPlots}
  \end{center}
\end{figure}

The following decays, and their charge conjugates, are identified:
 K$^{0}_{\rm S} \to \pi^{+} \pi^{-}$, $\Lambda \to $p$ \pi^{-}$, 
D$^{*+} \to$ D$^{0}$(K$^{-}\pi^{+})\pi^{+}$.
This ensemble of final states provides a complete set of charged particle types
 needed to comprehensively assess the RICH detectors hadron PID performance. 
As demonstrated in Fig.~\ref{fig:MassPlots}, the K$^{0}_{\rm S}$, $\Lambda$, and
  D$^{*}$ selections have extremely high purity. 

While high purity samples of the control modes can be gathered through purely 
kinematic requirements alone, the residual backgrounds present within each 
must still be accounted for. 
To distinguish background from signal, a likelihood
technique, called  $\!\!{}_{\phantom{1}s}\mathcal{P}lot$ 
\cite{Pivk:2004ty}, 
 is used, where the invariant mass of the composite particle
K$^0_{\rm S}, \Lambda$, D$^{0}$ is used as the discriminating variable.

\begin{figure}[h]
  \begin{center}
    \subfigure[]{\label{fig:Pion_2D}
      \includegraphics[width=0.4 \textwidth]{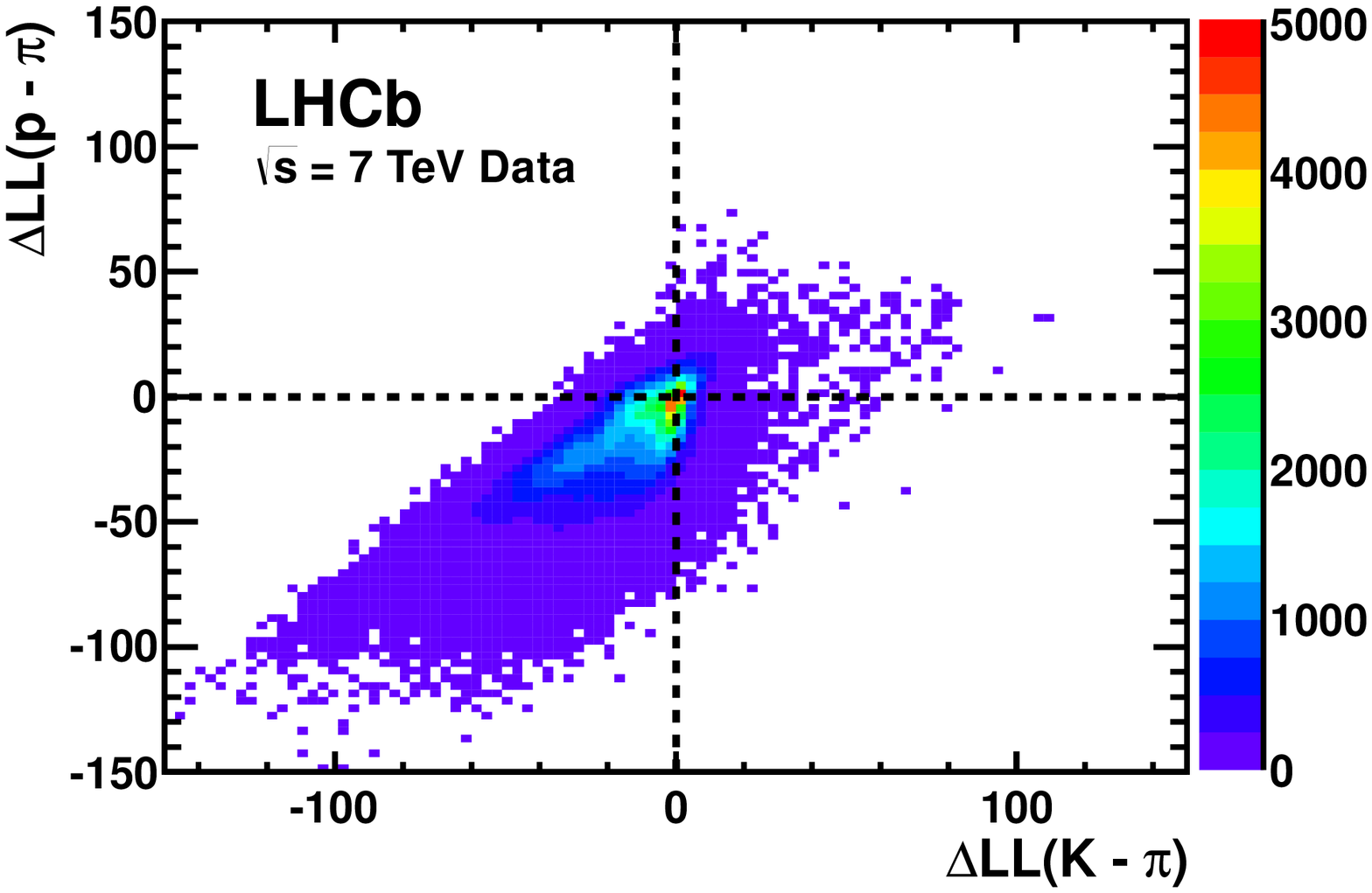}}
    \subfigure[]{\label{fig:Kaon_2D}
      \includegraphics[width=0.4 \textwidth]{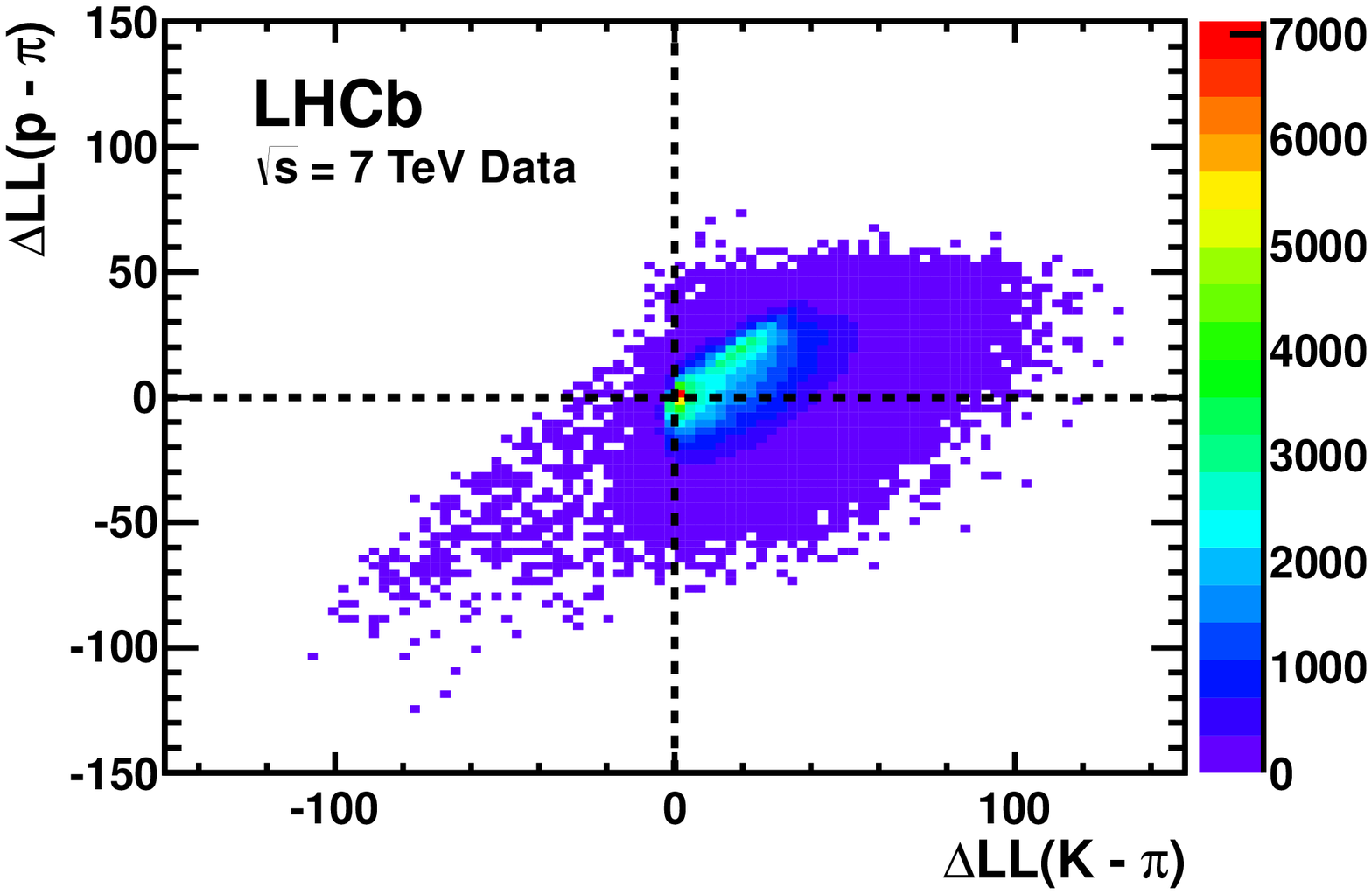}}
    \subfigure[]{\label{fig:Proton_2D}
      \includegraphics[width=0.4 \textwidth]{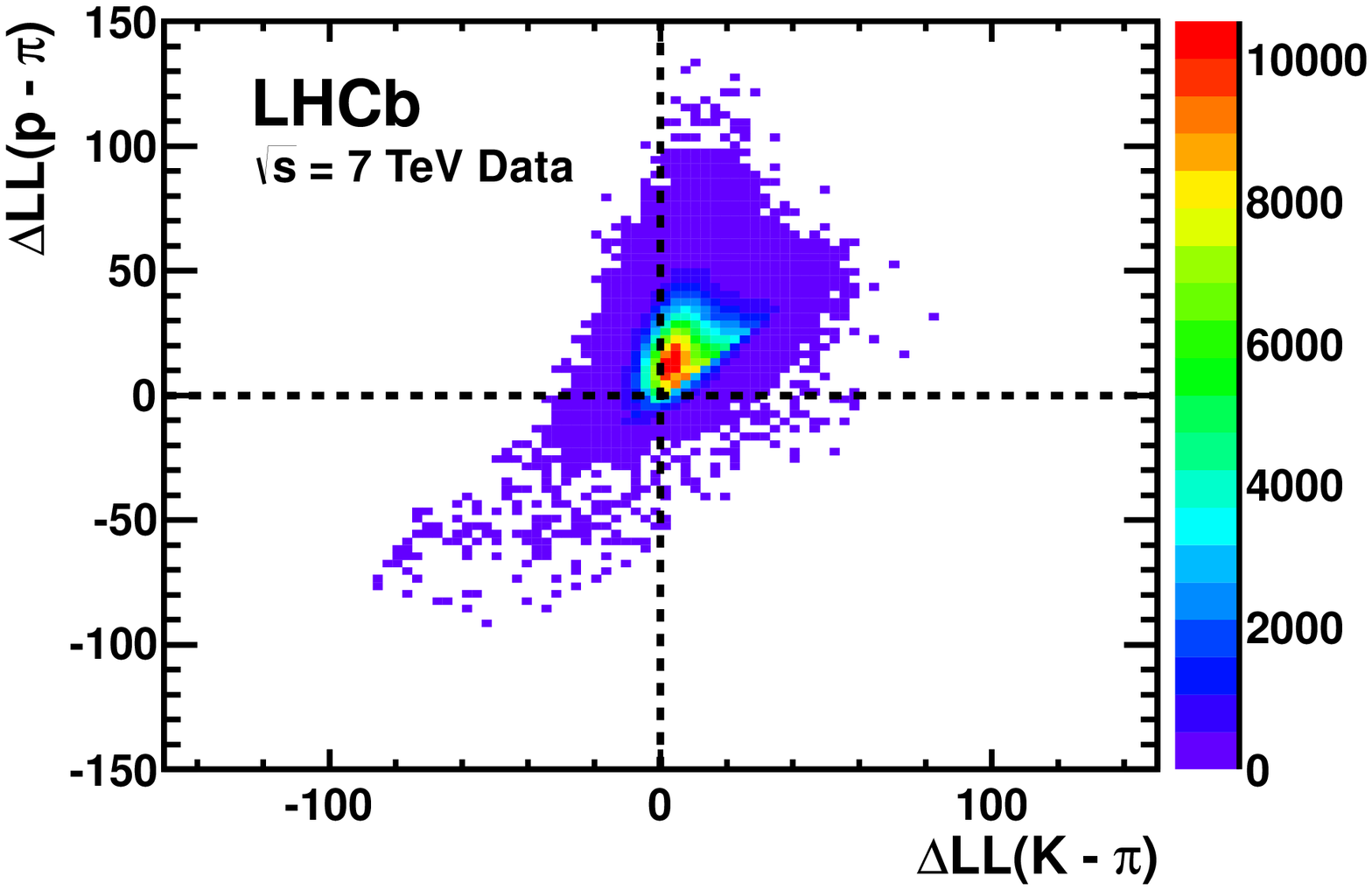}}
    \caption{Distribution of $\rm\Delta log \mathcal{L}(K - \pi)$ against $\rm\Delta
   log \mathcal{L}(p - \pi)$ for (a) pions, (b) kaons and (c) protons extracted from the control samples}
    \label{fig:DLLs_2D}
  \end{center}
\end{figure}

The power of the RICH PID can be appreciated by considering the
 $\rm \Delta log \mathcal{L}$ distributions for each track type from the 
control samples. Figures~\ref{fig:DLLs_2D}(a-c) show the 
corresponding distributions in the 2D plane of
 $\rm \Delta log \mathcal{L}(K -\pi)$ versus 
$\rm \Delta log \mathcal{L}(p - \pi)$. 
Each particle type is seen within a quadrant of the two dimensional
 $\rm \Delta log \mathcal{L}$ space, and demonstrates
 the powerful discrimination of the RICH.

\subsection{PID performance}
\label{sec:PID}

Utilizing the log-likelihood values obtained from the control channels,
 one is able to study the discrimination achievable between any pair of track 
types by
 imposing requirements on their differences, such as $\Delta{\rm log}(K -\pi)$.
 Figure 17 demonstrates the kaon efficiency (kaons identified as kaons) and 
pion misidentification (pions misidentified as kaons), as a function of 
particle momentum, obtained from imposing two different requirements on this 
distribution. 
 Requiring that the likelihood for each track with the kaon mass 
hypothesis be larger than that with the pion hypothesis, i.e.
 $\rm\Delta log \mathcal{L}(K-\pi)>0$, 
and averaging over the momentum range 2 - 100 GeV/$c$, the kaon efficiency
 and pion misidentification fraction are found to be 
$\sim 95\%$ and $\sim 10\%$, respectively. 
The alternative PID requirement of $\rm\Delta log \mathcal{L}(K-\pi)>5$
 illustrates that the misidentification rate can be significantly reduced to 
$\sim 3\%$ for a kaon efficiency of $\sim 85\%$. 
Figure~\ref{fig:MCKPiSeparation} shows the corresponding efficiencies and
 misidentification fractions in simulation.
In addition to K/$\pi$ separation, both p/$\pi$ and p/K separation are equally 
vital for a large number of physics analyses at LHCb. 
Figure~\ref{fig:PpiSeparation}
demonstrates the 
separation achievable between protons and pions when imposing the PID 
requirements 
$\rm\Delta\mathcal{L}(p -\pi)>0$ and $\rm\Delta\mathcal{L}(p -\pi)>5$.
Finally, Fig.~\ref{fig:PKSeparation}
shows the discrimination achievable between protons and 
kaons when imposing the requirements  $\rm\Delta\mathcal{L}(p - K)>0$ and 
$\rm\Delta\mathcal{L}(p - K)>5$. 

\begin{figure}
\begin{center}
  \includegraphics[width=0.58\textwidth]{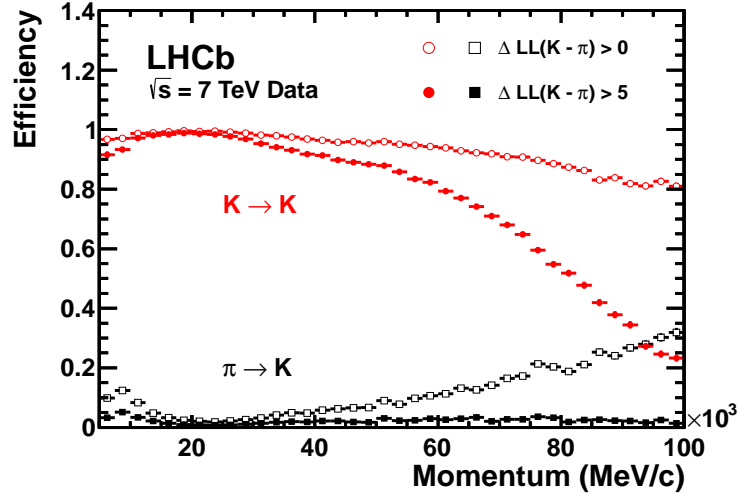}
  \caption{Kaon identification efficiency and pion misidentification rate
  measured on data as a function of track momentum. Two different $\rm\Delta log
  \mathcal{L}(K-\pi)$ requirements have been imposed on the samples, resulting
  in the open and filled marker distributions, respectively }

  \label{fig:KPiSeparation}
\end{center}
\end{figure}

\begin{figure}
\begin{center}
  \includegraphics[width=0.58\textwidth]{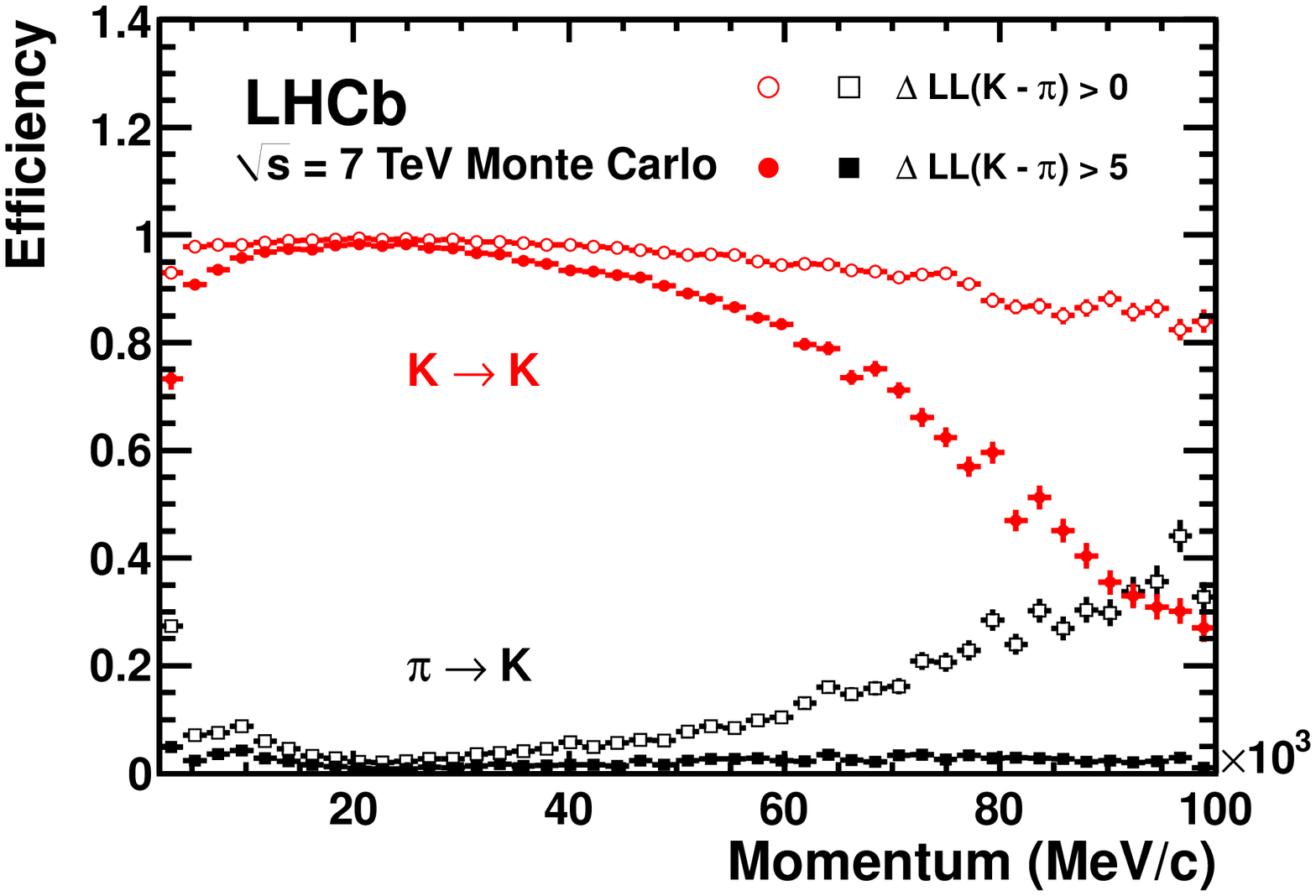}
  \caption{Kaon identification efficiency and pion misidentification rate
  measured using simulated events as a function of track momentum. Two different
  $\rm\Delta log \mathcal{L}(K-\pi)$ requirements have been imposed on the
  samples, resulting in the open and filled marker distributions,
  respectively }

  \label{fig:MCKPiSeparation}
\end{center}
\end{figure}

\begin{figure}
\begin{center}
  \includegraphics[width=0.58\textwidth]{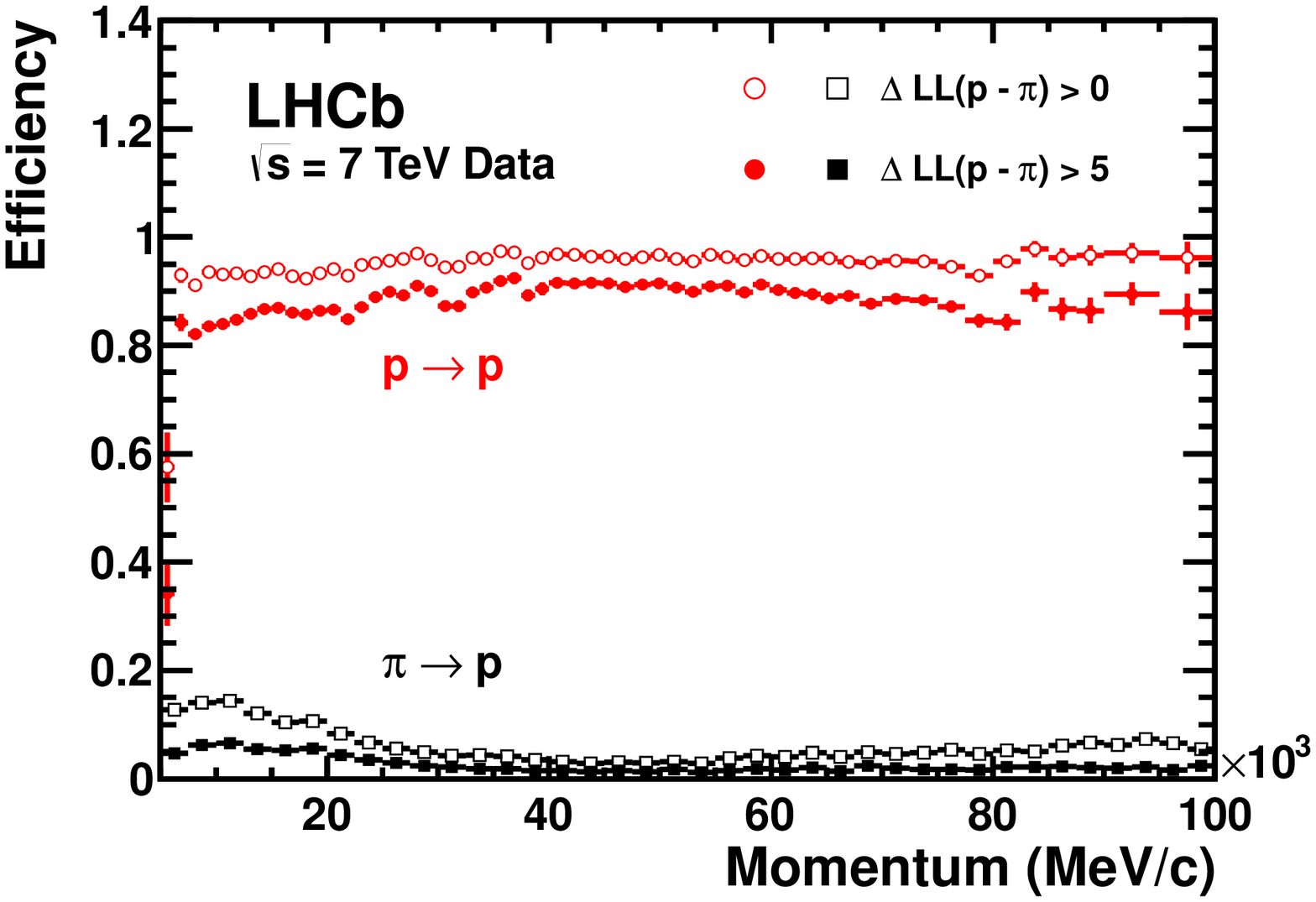}
  \caption{Proton identification efficiency and pion misidentification rate
  measured on data as a function of track momentum.
Two different $\rm\Delta log
  \mathcal{L}(p-\pi)$ requirements have been imposed on the samples, resulting
  in the open and filled marker distributions, respectively }

  \label{fig:PpiSeparation}
\end{center}
\end{figure}

\begin{figure}
\begin{center}
  \includegraphics[width=0.58\textwidth]{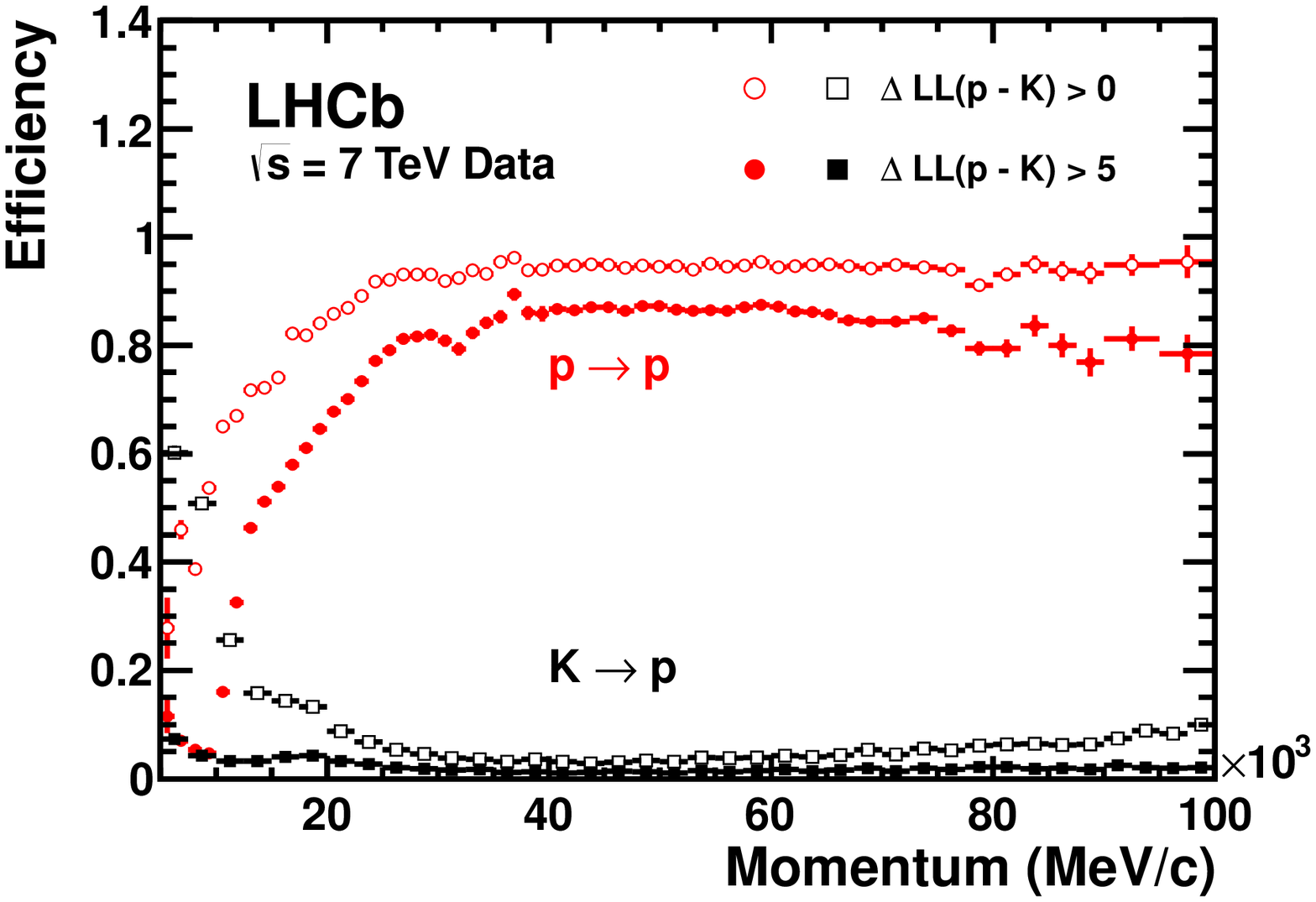}
  \caption{Proton identification efficiency and kaon misidentification rate
  measured on data as a function of track momentum.
Two different $\rm\Delta log
  \mathcal{L}(p-K)$ requirements have been imposed on the samples, resulting
  in the open and filled marker distributions, respectively }

  \label{fig:PKSeparation}
\end{center}
\end{figure}

Together, 
Figures~\ref{fig:KPiSeparation},
~\ref{fig:PpiSeparation} and
~\ref{fig:PKSeparation}
 demonstrate the
 RICH detectors ability to discriminate any pair of track types, from the set
 of kaons, pions and protons, albeit for the PID requirements quoted.

\subsection{Performance as a function of event multiplicity}
The current running conditions\footnote{The LHCb RICH detector was designed to 
run with 0.6 interaction per 
bunch crossing. However the current operating conditions have 1.6 interactions 
per bunch crossing. },
 with increased particle multiplicities, provide an insightful glimpse of the 
RICH performance at high luminosity running.

\begin{figure}[ht]
  \begin{center}
    \subfigure[]{\label{fig:Perf:nTrack}
      \includegraphics[width=0.45 \textwidth]{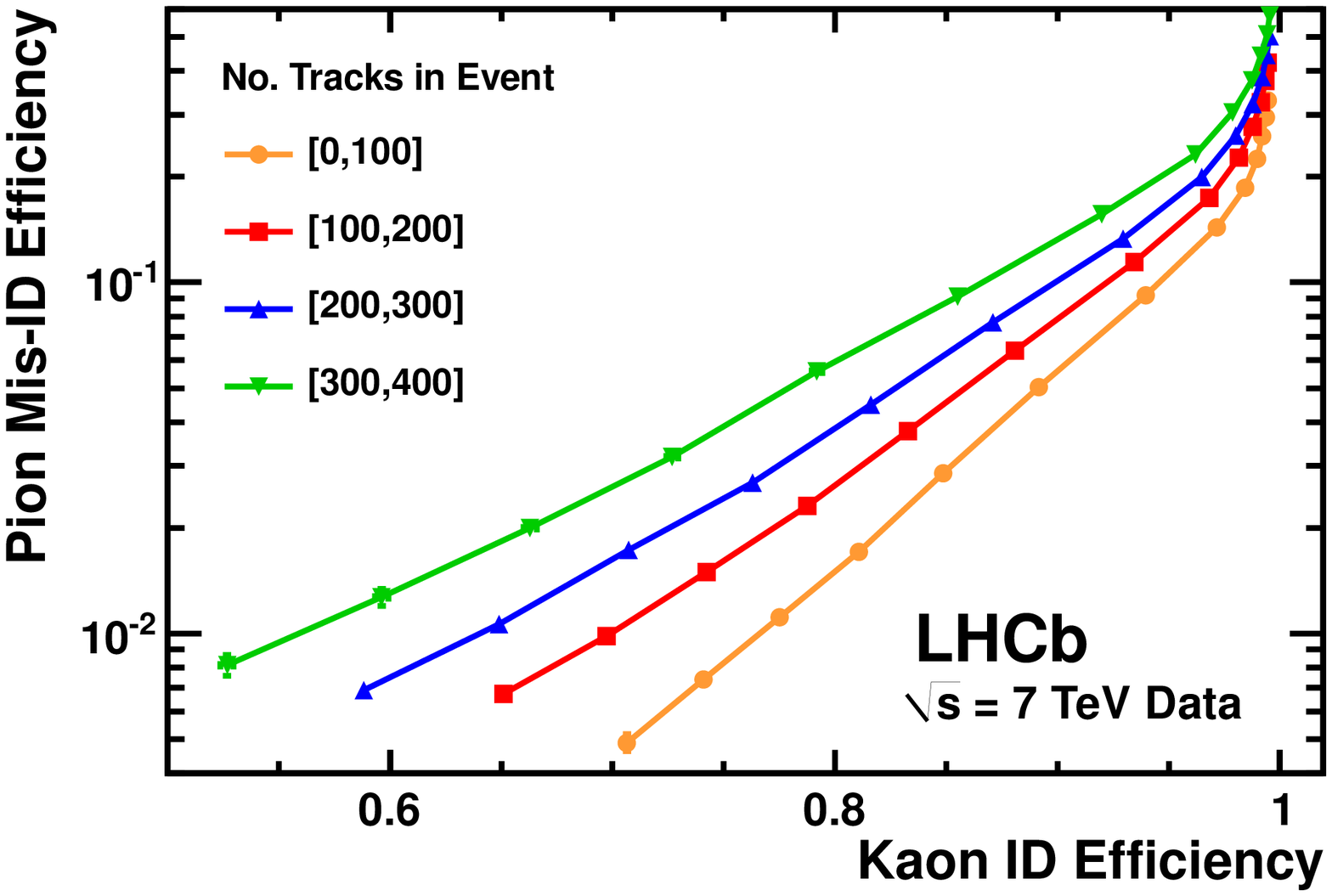}}
    \subfigure[]{\label{fig:Perf:nPV}
      \includegraphics[width=0.45 \textwidth]{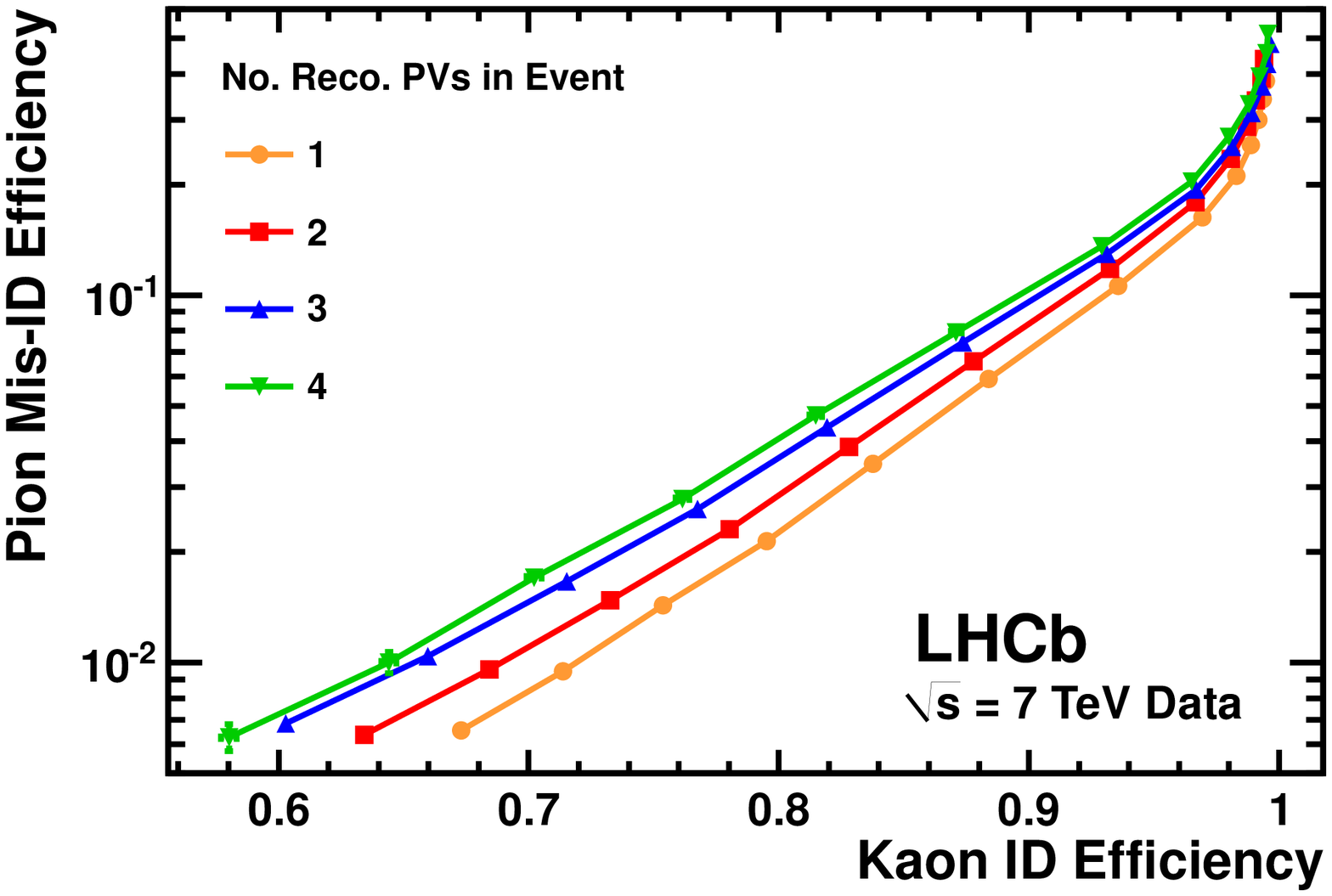}}
    \caption{Pion misidentification fraction versus kaon identification
    efficiency as measured in 7~TeV LHCb collisions: (a) as a function of track
    multiplicity, and (b) as a function of the number of reconstructed primary
    vertices. The efficiencies are averaged over all particle momenta
}
    \label{fig:Perf:Multiplicity}
  \end{center}
\end{figure}

Figure~\ref{fig:Perf:Multiplicity} shows the pion misidentification fraction 
versus the kaon identification efficiency as a function of (a) track 
multiplicity and (b) the number of reconstructed primary vertices, as the 
requirement on the likelihood difference $\rm \Delta log \mathcal{L}(K-\pi)$ 
is varied. The results demonstrate, as expected, some degradation in PID 
performance with increased interaction multiplicity. The $K/\pi$ separation is,
 however, extremely robust right up to the highest interaction multiplicities 
and thus gives confidence that the current RICH system is suitable for 
operation at the higher luminosities foreseen in the future.

\section{Conclusions}

The RICH detector was designed specifically for the physics program of LHCb. 
It has been in operation since the end of 2009. 
The RICH detector has operated with high efficiency during these first 
three years of LHC running.  It has demonstrated a PID 
performance that is well up to design specifications and that allows the 
extraction of physics results in all sectors of $b$ and $c$ quark decays, 
 in particular
 of the rare phenomena  which may allow the discovery of new physics at the LHC.

The performance of the RICH particle identification has been studied
 with the LHC collisions taken since the startup of the LHC machine. 
Studies of 
the decays of K$^0$, $\Lambda^0$ and D$^*$ provide a source of $\pi$, K, p
 identified kinematically for which the RICH identification performance can be
 established. The precise alignment and calibration procedures are crucial to
 reach the designed performance. The Cherenkov angle resolutions are in good 
agreement with the expected design performance for the gas radiators, and are 
still being improved for the aerogel radiator.

 \section*{Acknowledgements}

This complex detector could only be constructed with the dedicated effort of 
many technical collaborators in the institutes forming the LHCb RICH 
collaboration. 
A special acknowledgement goes to all our LHCb collaborators who over the
years have contributed to obtain the results presented in this paper.
\noindent We express our gratitude to our colleagues in the CERN
accelerator departments for the excellent performance of the LHC. We
thank the technical and administrative staff at the LHCb
institutes. We acknowledge support from CERN and from the national
agencies: CAPES, CNPq, FAPERJ and FINEP (Brazil); NSFC (China);
CNRS/IN2P3 and Region Auvergne (France); BMBF, DFG, HGF and MPG
(Germany); SFI (Ireland); INFN (Italy); FOM and NWO (The Netherlands);
SCSR (Poland); ANCS/IFA (Romania); MinES, Rosatom, RFBR and NRC
``Kurchatov Institute'' (Russia); MinECo, XuntaGal and GENCAT (Spain);
SNSF and SER (Switzerland); NAS Ukraine (Ukraine); STFC (United
Kingdom); NSF (USA). We also acknowledge the support received from the
ERC under FP7. The Tier1 computing centres are supported by IN2P3
(France), KIT and BMBF (Germany), INFN (Italy), NWO and SURF (The
Netherlands), PIC (Spain), GridPP (United Kingdom). We are thankful
for the computing resources put at our disposal by Yandex LLC
(Russia), as well as to the communities behind the multiple open
source software packages that we depend on.






\begin{thebibliography}{99}


\bibitem{Detector} LHCb collaboration, A.A.\ Alves Jr.\ {\em et al.},
 JINST 3 (2008) S08005  

\bibitem{results} see for ex.: LHCb Collaboration, R.\ Aaij {\em et al.},\\ 
 Phys. Rev. Lett. {\bf 108} (2012) 251802,\\
 Phys. Lett. B {\bf 712} (2012) 203,\\
 Phys. Rev. D {\bf 85} (2012) 112013,\\
 Phys. Rev. Lett. {\bf 108} (2012) 201601 

\bibitem{vagnoni} LHCb collaboration, R.\ Aaij\ {\em et al.}, 
JHEP {\bf 10} (2012) 037   

\bibitem{tagging} LHCb collaboration, R.\ Aaij\ {\em et al.},   
Eur. Phys. J. C {\bf 72} (2012) 2022

\bibitem{trigger} LHCb collaboration, R.A.\ Nobrega\ {\em et al.},   
CERN/LHCC 2003-031, LHCb TDR 10 2003 

\bibitem{scintillation}J.E. Hesser and K. Dressler, The Journal of Chemical 
Physics 47 (1967) 3443  


\bibitem{Aerogel} T.\ Bellunato {\em et al.}, 
Nucl.\ Inst.\ Meth.\ A {\bf 527} (2004) 319

\bibitem{HPD} M.\ Alemi {\em et al.}, Nucl.\ Inst.\ Meth.\ A {\bf 449} (2000) 48



\bibitem{lownoise} S. Eisenhardt, Nucl.\ Inst.\ Meth.\ A {\bf 595} (2008) 142\\
M. Moritz {\em et al.}, {\em Performance Study of New Pixel Hybrid Photon 
Detector Prototypes for the LHCb RICH Counters}, IEEE TRANSACTIONS ON NUCLEAR 
SCIENCE, VOL. 51, NO. 3, JUNE 2004


\bibitem{RichDCS} A. Papanestis {\em The LHCb RICH Detector Control System},
Proceedings of ICALEPCS2009 page 340 \\
F. Fontanelli {\em et al.,}Nucl.\ Inst.\ Meth.\ A {\bf 604} (2009) 675\\
C. Arnaboldi {\em et al.,}NSS'07 \ IEEE\ Volume 1 (2007) 677


\bibitem{JCOP} Holme, M. Gonzalez Berges, P. Golonka and S. Schmeling, 
{\em The JCOP framework}, Conf.Proc.C051010:WE2 (2005) 60   

\bibitem{CondDB} M Clemencic, J. Phys. Conf. Ser. 119 (2008) 072010





\bibitem{adinolfi} M. Adinolfi {\em et al.}, Nucl.\ Inst.\ Meth.\ A {\bf 572} (2007) 689

\bibitem{MDMSyst} A. Borgia {\em et al.}, arXiv:1206.0253, submitted to Nucl.\ Inst.\ Meth.\ A

\bibitem{bi:Thesis} R. Cardinale {\em et al.}, JINST {\bf 6} (2011) P06010 

\bibitem{bi:TGnote} T. Gys, {\em Magnetic field simulations for 
the LHCb RICH-2 detector}, Note LHCb-2002-029

\bibitem{Alignment} W. Baldini {\em et al.,}, {\em LHCb alignment strategy}, 
Note LHCb-2006-035

\bibitem{GaudiA} G. Barrand {\em et al.}, {\em
 GAUDI - A software architecture and framework for building HEP data
  processing applications}, Comput.Phys.Commun.140:45-55,2001

\bibitem{GaudiB} M. Clemencic {\em et al.},
{\em Recent developments in the LHCb software framework Gaudi},
J.Phys.Conf.Ser.219:042006,2010  

\bibitem{forty} R. Forty,
Nucl.\ Inst.\ Meth.\ A {\bf 384} (1996) 167

\bibitem{geant4} S. Agostinelli {\em et al.},
  Nucl.\ Inst.\ Meth.\  A {\bf 506} (2003) 250

S. Easo {\em et al.}, 
{\em Simulation of LHCb RICH detectors using GEANT4},
 IEEE Trans. Nucl. Sci. {\bf 52} (2005) 1665

\bibitem{nnet} S. Gorbunov and I. Kisel,
CBM-SOFT-note-2005-002


\bibitem{Pivk:2004ty}
  M.~Pivk and F.~R.~Le Diberder,
  Nucl.\ Inst.\ Meth.\  A {\bf 555} (2005) 356



\end{thebibliography}
\end{document}